\documentclass[a4paper,12pt]{iopart}
\usepackage{graphicx}
\usepackage[T1]{fontenc} 
\usepackage{epstopdf}
\usepackage{bm,amsfonts,lineno,hyperref,amssymb,microtype,float,amsgen,amsbsy}
\usepackage{color}
\usepackage{array}
\usepackage{iopams}
\usepackage{graphicx}
\usepackage{hyperref}
\usepackage{amstext}
\usepackage{savesym}
 \expandafter\let\csname equation*\endcsname\relax
\expandafter\let\csname endequation*\endcsname\relax
\usepackage{amsmath}
\usepackage{txfonts}
\usepackage{microtype}
\usepackage[a4paper, total={7in, 9.1in}]{geometry}

\begin{document}
	
	\title{Traversable Lorentzian wormhole on the Shtanov-Sahni braneworld with matter obeying the energy conditions}
	
\author{Rikpratik Sengupta$^1$, Shounak Ghosh$^2$, Mehedi Kalam$^1$}

\address{$^1$ Department of Physics, Aliah University, Kolkata 700160, West Bengal, India}
\address{$^2$ Department of Physics, Indian Institute of Engineering Science and Technology, Shibpur, Howrah 711103, West Bengal, India}

\ead{$^1$ rikpratik.sengupta@gmail.com}
\ead{$^2$ shounak.rs2015@physics.iiests.ac.in}
\ead{$^1$ kalam@associates.iucaa.in}

	\date{\today}
	
	\begin{abstract}
		In this paper we have explored the possibility of constructing a traversable wormhole on the Shtanov-Sahni braneworld with a timelike extra dimension. We find that the Weyl curvature singularity at the throat of the wormhole can be removed with physical matter satisfying the NEC $\rho+p \geq 0$, even in the absence of any effective $\Lambda$-term or any type of charge source on the brane. (The NEC is however violated by the effective matter description on the brane arising due to effects of higher dimensional gravity.) Besides satisfying NEC the matter constituting the wormhole also satisfies the Strong Energy Condition (SEC), $\rho+3p \geq 0$, leading to the interesting possibility that normal matter on the brane may be harnessed into a wormhole. Incidentally, these conditions also need to be satisfied to realize a non-singular bounce and cyclic cosmology on the brane~\cite{Sahni4} where both past and future singularities can be averted. Thus, such a cyclic universe on the brane, constituted of normal matter can naturally  contain wormholes. The wormhole shape function on the brane with a time-like extra dimension represents the tubular structure of the wormhole spreading out at large radial distances much better than in wormholes constructed in a braneworld with a spacelike extra dimension and have considerably lower mass resulting in minimization of the amount of matter required to construct a wormhole. Wormholes in the Shtanov-Sahni (SS) braneworld also have sufficiently low tidal forces, facilitating traversability. Additionally they are found to be stable and exhibit a repulsive geometry. We are left with the intriguing possibility that both types of curvature singularity can be resolved with the SS model, which we discuss at the end of the concluding section.
	\end{abstract}
	
	\maketitle

	\section{INTRODUCTION}
	
	Wormholes are a class of topological object with a trivial boundary and non-simply-connected interior, which first appeared as a solution to the field equations of Einstein's General Relativity (GR) in 1916~\cite{Ludwig1916}. Later on, Einstein himself along with Rosen extended the idea to speculate the possible existence of bridges connecting two different spacetime regions. Such bridges were believed to create a shortcut through spacetime, resulting in a reduced travel distance and time~\cite{ER1935}. The detailed mathematical structure of a wormhole was investigated few decades later by Fuller and Wheeler~\cite{FH1962} for the Schwarzschild geometry, resulting in a tubular structure spreading out to be asymptotically flat. The two different spacetime regions were connected by the throat of the wormhole.  However, it was found in their analysis that such a structure  would collapse at the throat from the instability resulting due to the gravitational attraction exerted by matter present in the two different spacetime regions in the opposite directions. As a consequence, the throat would pinch-off, resulting in possible appearance of a Weyl singularity at the throat resulting from infinitely large tidal forces. So, such objects remained of very little physical interest until Morris and Thorne~\cite{MT1988} proposed a generalized prescription to possibly avoid the pinch-off at the throat.
	
	They prescribed certain properties for one of the metric potentials termed as the shape function, which represents the shape of the wormhole. According to their prescription, the shape function at the throat radius must be equal to the throat radius. For all radial distances within the surface of the wormhole greater than the throat radius, the shape function at a particular radial distance must be less than the radial distance itself.  Most importantly, the derivative of the shape function with respect to the radial distance at the throat radius must be less than unity. This implies a violation of the null energy condition (NEC) at the throat. This is known as the 'flare-out' condition. Additionally, the surface density and surface pressure must vanish at the boundary of the wormhole.
	
	The motivation for the main component of their prescription can be analyzed from two points of view. Physically, if some 'exotic' matter is introduced at the throat with negative pressure, such that it is gravitationally repulsive in nature, then the instability arising from attraction due to matter in the two different spacetime regions can be overcome as they will be repelled by the exotic matter due to its peculiar property. Thus, the throat will remain stable and 'flare-out' keeping the bridge-like structure stable. From a mathematical point of view, a Weyl type of curvature singularity, which is indeed a physical singularity indicating geodesic incompleteness, can be averted even in a relativistic context by the violation of the null energy condition (NEC), as inferred from the singularity theorems due to Penrose and Hawking~\cite{Hawking}. It is interesting to note that matter sources violating the strong energy condition (SEC) and sometimes also the NEC, play an important role in cosmology~\cite{Sahni1,Peebles2003,LudwickMpla} to explain the current accelerating state~\cite{Riess1998,Perlmutter1999} of the universe.
	
	In order to explain the current accelerating phase of the universe, the Einstein field equations (EFE) have to be modified. This is possible either by modifying the geometry or the matter sector. In order to modify the geometry sector, the standard Einstein-Hilbert (EH) action has to replaced such that the EH action can be recovered for suitable choice of the model parameters~\cite{Harko,NojiriR,Sahni2,DDG}. On the other hand, the modification in the matter sector~\cite{Sahni3,Sahni1,Peebles2003,LudwickMpla,Sengupta4} arises from introduction of a variety of matter sources violating one or more of the energy conditions, ranging from the well known Einstein's Cosmological Constant to scalar fields with a negative kinetic energy term dubbed as the 'phantom'. The phantom EoS with the EoS parameter being less than -1 can however be realized effectively as a modification in the geometry sector arising from the higher dimensional braneworld scenario, without considering any such exotic matter source~\cite{Sahni2}. Also, it is now well accepted that standard GR fails to work adequately at spacetime regions involving diverging energy densities and spacetime curvatures. The divergences are believed to occur due to the shortcomings of GR and for explaining such situations, GR needs to be replaced a modified gravity theory that reduce to standard GR at low energy limits. The two main contenders in this aspect are the higher dimensional braneworld models~\cite{Randall1,Randall2,Sahni2,Sahni4}, inspired from Superstring/M-theories~\cite{Polchinski1,Polchinski2,Gasperini} and an alternative four dimensional approach known as the Loop quantum gravity (LQG)~\cite{Rovelli,Bojowald}.
	
	The braneworld models have attracted a lot of attention in recent times. One of the first braneworld models that became extremely popular, was proposed by Randall and Sundrum (RS) in 1999~\cite{Randall1,Randall2} as an attempt to solve the hierarchy problem~\cite{VSK} in particle physics. The comparative weakness of gravity as compared to the electroweak forces could be explained by assuming our universe to be contained in a (3+1)- dimensional membrane embedded in a higher dimensional bulk spacetime. The standard model fields of particle physics are considered to be confined to the brane, while gravity can access the bulk spacetime which is essentially AntideSitter ($AdS_5$). The negative cosmological constant on the brane confines the gravity close to the brane. Their first model consisted of two parallel braneworlds, but in the second model~\cite{Randall2} the brane with a negative tension was moved to infinity. The RS single brane model has a spacelike extra dimension and finds wide range of application in both cosmology~\cite{Binetruy,Maeda,Langlois,Chen,Kiritsis,Campos,Sengupta1,Maartens,Sengupta} and astrophysics~\cite{Wiseman2,Germani,Deruelle,Wiseman,Visser,Creek,Pal,Bruni,Govender,Sengupta5}. Soon after, a dual to the RS-II single brane model was proposed by Shtanov and Sahni (SS), with a timelike extra dimension replacing the spacelike extra dimension~\cite{Sahni4}. For a spacelike extra dimension the bulk space has a Lorentzian signature but, for the later, the bulk space has a signature (-,-,+,+,+). Such a braneworld is characterized by a negative brane tension and positive bulk cosmological constant, contrary to the RS-II braneworld. The most interesting feature of this model is that the contracting phase realizes a natural bounce, avoiding reaching a singular state, even if matter on the brane satisfies the energy conditions appropriately. Nothing prevents the non-singular bounce from happening an infinite number of times and in the presence of a massive scalar field, the amplitude increases for each successive cycle leading to a resolution of the flatness problem~\cite{S6,S7,S8}. A spatially closed universe consisting of matter satisfying the strong energy condition (SEC) can undergo the cyclic behaviour and in general a universe consisting of matter satisfying the null energy condition (NEC) can transit through a non-singular bounce both in the past and future, provided our universe is considered to be a (3+1)-dimensional brane embedded in the bulk with timelike extra dimension. There is an issue regarding the tachyonic nature of the Kaluza-Klein gravitational modes in a model with time-like extra dimension as discussed in~\cite{Sahni4} and has been addressed in ~\cite{S10}.
	
	The possibility of wormhole formation have previously been investigated modifying both the matter~\cite{Barcelo,Hayward,Picon,Sushkov,Lobo,Zaslavskii,Chakraborty,Sengupta3} as well geometry~\cite{Bhawal,Bhadra,Eiroa,Bertolami,Moraes,Agnese,He,Dzhunushaliev,Bronnikov,Lobo2,Banerjee,Chakraborty2,Rahaman1,Wang,Sengupta2} sectors. Wormholes have also been explored in the framework of extra dimensions in the context of the RS II braneworld model~\cite{Sengupta2}. Generally to realize traversable wormholes in a relativistic context phantom matter is required to violate the NEC in order to make the throat stable but in a recent work~\cite{Salcedo}, it has been showed that using $Z_2$-reflection symmetry at the throat, a wormhole can be constructed with normal matter containing coupled Maxwell and Dirac fields but at the expense of coexisting particles and antiparticles which do not annihilate and non-smooth geometry and matter sectors. This has been solved by Konoplya and Zhidenko~\cite{Konoplya} by not considering the $Z_2$ symmetry and the problem of particle-antiparticle coexistence at the throat can be resolved besides obtaining a smooth metric and matter field. The braneworld model considered by us however invokes the $Z_2$ symmetry but not in a relativistic context or specifically at the wormhole throat and is a generic feature of the model.
	
	\textbf{{There have been a few attempts to design ideas of possible detection or observation of wormholes in literature~\cite{Shaikh,Li,Ohgami,Tsukamoto,Nandi,Shaikh2}. One such possibility arises as a consequence of the individual fluxes associated with the two different spacetimes connected by the wormhole not being conserved. The effect of this can be realised by mutual effects on massive objects in the vicinity of either wormhole mouths. An example of this may be such unexplained effect on the orbit of stars which are present in the vicinity of the black hole at the centre of our galaxy that is believed to harbor a possible wormhole~\cite{DS}. The mass density of the wormhole has been constrained by obtaining an upper limit by studying its possible micro-lensing effects which are believed to resemble gamma ray bursts~\cite{Torres}. There also arises a possibility of emission of radiation pulses due to a wormhole which can possible be detected~\cite{D}. The quasinormal black hole ringing can be distinguished from a wormhole supported by phantom matter with a specific EoS. With or without applying the thin-shell formalism, the ringing can be distinguished either at late times or at all times~\cite{KZ}.}}.	
	
	In this paper, we attempt to explore the possibility of existence of wormholes in the context of the braneworld model with a timelike extra dimension, since it has many interesting additional features absent in the RS-II models, specially avoiding the cosmological singularities and realizing the possibility of a cyclic universe scenario. In the following section, we solve the EFE for the wormhole metric on the brane to obtain the unknown metric potential and study the validity of the NEC. The Israel-Darmois junction conditions~\cite{Israel,Darmois} are formulated at the surface of the wormhole to evaluate the unknown model parameters. Finally, the tidal acceleration is obtained at the throat to ensure traversability besides checking on the stability involving a linearized stability analysis along with computation of the surface redshift and commenting on the nature of the wormhole.
	\section{Mathematical Model of the Brane Wormhole}
	
	In this section we attempt to construct a mathematical model of a spherically symmetric, static, traversable wormhole on the brane
	with a timelike extra dimension, that is stable under linear stability analysis. We check the various criteria that must be satisfied
	by the wormhole solution to be rendered as physically realistic.
	
	\subsection{Modified EFE on the Brane}
	
	A spherically symmetric and static line element has the well known form
	\begin{equation}
		ds^2=-e^{\nu(r)}dt^2+e^{\lambda(r)}dr^2+r^2(d\theta^2+sin^2\theta d\phi^2).
	\end{equation}
	
	There is modification of the EH action representing the geometry sector for the braneworld resulting in the appearance of additional terms in the EFE. The modified EFE on the brane may be written in the most general form as~\cite{Sahni5}
	\begin{equation}
		m^2 G_{\mu \nu}+\sigma h_{\mu \nu}= T_{\mu \nu}+ \epsilon M^3(K_{\mu \nu}-h_{\mu \nu}K),
	\end{equation}
	Making use of the Gauss identity on the brane, the modified EFE reads~\cite{Sahni5}
	\begin{equation}
		G_{\mu \nu}+\Lambda_{eff}h_{\mu \nu}=8\pi G_{eff} T_{\mu \nu}+\frac{\epsilon}{1+\beta}\bigg[\frac{S_{\mu \nu}}{M^6}-W_{\mu \nu}\bigg]
	\end{equation}
	The parameter $\beta$ is given by $\beta=\frac{2\sigma m^2}{3M^6}$, where $m$ is the Planck mass in four dimensions, $M$ is the five dimensional Planck mass, $\sigma$ denotes the brane tension,
	$G_{\mu \nu}$ is the Einstein tensor on the (3+1)-dimensional brane and $T_{\mu \nu}$ is the energy momentum tensor on the brane.
	Here $K_{\mu \nu}$ is the extrinsic curvature and $K$ represents its trace;
	$h_{\mu \nu}=g_{\mu \nu}-\epsilon n_{\mu}n_{\nu}$ is the induced metric on the brane, such that $n^{\mu}$ represents the vector
	field of the inner unit normal to the brane and $g_{\mu \nu}$ is the metric on the bulk.  The parameter $\epsilon=\pm 1$ determines the signature of the bulk space.
	If we choose $\epsilon=1$ then the bulk spacetime has a Lorentzian signature and the extra dimension is spacelike. We get the
	RS II model in this case. On the other hand, a choice $\epsilon=-1$ deviates the bulk signature from being Lorentzian and the
	extra dimension is timelike giving the SS braneworld model. The effective cosmological constant on the brane is given by $\Lambda_{eff}=\frac{\Lambda_{RS}}{1+\beta}$, where $\Lambda_{RS}$ is the effective cosmological constant on the RS II brane and the effective gravitational constant on the brane is given by $8\pi G_{eff}=\frac{\beta}{m^2(1+\beta)}$. The terms within the square brackets denote the correction terms which arise as a consequence of the extra dimensional effects. The term $S_{\mu \nu}$ turns out to be quadratic in stress-energy and is computed from the bare Einstein equation $E_{\mu \nu}=m^2 G_{\mu \nu}-T_{\mu \nu}$ obtained from Eq. (2). It is given by
	\begin{equation}
		S_{\mu \nu}=\frac{1}{3}EE_{\mu \nu}-E_{\mu \alpha}E^{\alpha}_{\nu}+\frac{1}{2}\bigg(E_{\alpha \gamma}E^{\alpha \gamma}-\frac{1}{3}E^2\bigg)h_{\mu \nu}
	\end{equation}
	The other term $W_{\mu \nu}$ is the projection of the bulk Weyl tensor $W_{\mu \nu \alpha \gamma}$ on the brane that is responsible for the corrections encoding the bulk graviton effects. It is defined as $W_{\mu \nu}=n^{\alpha} n^{\gamma} W_{\mu \nu \alpha \gamma}$.
	The Weyl projection appears in Eq. (3) as the stess-energy tensor of some additional matter, resulting in the introduction of extra effective energy density and effective pressure terms in the modified field equations. The equation of state of the additional matter is that of radiation due to $W_{\mu \nu}$ being traceless. The correction terms are related to one another by the conservation equation
	\begin{equation}
		D^{\mu}\bigg(S_{\mu \nu}-M^6 W_{\mu \nu}\bigg)=0,	
	\end{equation}
	where $D^{\mu}$ denotes the covariant derivative on the brane associated with the induced metric $h_{\mu \nu}$. It is to be noted that we need not impose this equation additionally as it is a consequence of the presence of the projected Weyl term in the modified EFE on the brane.
	
	In this paper, we shall work in the RS limit where the induced curvature term on the brane is made to vanish by setting $m=0$. $m$ vanishes in the RS limit as the induced curvature (scalar curvature of the induced metric) on the brane generating from quantum correction to the matter action is not taken into account in the RS model.  The fundamental constant in the RS model is not the 4D Planck mass. The effective Planck mass on the brane is not equal to m but to a certain combination involving all constants, and it does not vanish in the case of $m = 0$. In fact, the RS model is a viable gravity theory on the brane even though it has $m = 0$ in our notation. Alternatively, a vanishing M indicates diminishing significance of extra dimensional effects and standard GR is recovered. It is worth mentioning that for a timelike extra dimension, a non-singular bounce can be obtained in the cosmological context irrespective of whether the induced curvature term is made to vanish or not. However in the presence of an induced curvature term in the action, there is an additional possibility of the universe commencing from a quasi-singular state that has the peculiar property of a non-divergent Hubble parameter and stress-energy while only the curvature tensor is divergent. The effective gravitational constant on the brane depends on the brane tension $\sigma$ and the parameter $\epsilon$ and may be written in the RS limit as $8 \pi G_{eff}=\frac{2\epsilon \sigma}{3 M^6}$. As mentioned, for the SS braneworld with a timelike extra dimension, $\epsilon=-1$. So, in order to make $G_{eff}$ positive, the brane tension $\sigma$ must be a negative quantity. Also, the effective cosmological
	constant on the brane in the $m=0$ limit is just a linear combination of the bulk cosmological constant term $\Lambda_5$ and the brane tension and is expressed as $\Lambda_{RS}=\frac{\Lambda_{5}}{2}+\frac{\epsilon \sigma^2}{3M^6}$. For the possibility of a vanishing effective lambda term on the brane, the bulk cosmological constant term must be positive as $\epsilon=-1$. Following RS, we assume $\Lambda_{RS}=0$ (implying a fine-tuning between the brane tension and bulk cosmological term) for constructing our wormhole model on the brane. As we consider the extra-dimension to be timelike, the evolution from the brane to the bulk is a Cauchy development, which is correctly posed and has a solution in the neighbourhood of the brane. So one can take an approach that one does not care what happens in the bulk ~\cite{S9}.
	
	The modified EFE on the brane for line element (1) assuming a perfect fluid source with isotropic pressure having stress-energy components $T_{\mu}^{\nu}=diag(-\rho,p,p,p)$ can be computed from Eqs. (2), (3) and (4) in geometrical units as
	\begin{eqnarray}
		e^{-\lambda}\left(\frac{\lambda^\prime}{r}-\frac{1}{r^2}\right)+\frac{1}{r^2}
		=8 \pi \left(\rho  \left( 1-\frac {\rho }{\rho_c} \right) \right)-\frac{12U}{\rho_c},\label{eq8}\\
		e^{-\lambda}\left(\frac{\nu^\prime}{r}+\frac{1}{r^2}\right) -\frac{1}{r^2}
		=8 \pi \left(p -{\frac {\rho  \left( p +\frac{\rho}{2} \right)}
			{\frac{\rho_c}{2}}}\right)-\frac{4U}{\rho_c}-\frac{8P}{\rho_c},\label{eq9}\\
		e^{-\lambda}\left(\frac{\nu''}{2}-\frac{\lambda^\prime \nu^\prime}{4}+\frac{{\nu^\prime}^2}{4}+\frac{\nu^\prime-\lambda^\prime}{2r}\right) = 8\pi \Bigg(p-{\frac{\rho \left(p+\frac{\rho}{2}\right)}{\frac{\rho_c}{2}}}\Bigg) -\frac{4U}{\rho_c}+\frac{4P}{\rho_c}.\label{eq10}
	\end{eqnarray}
	$\rho=\rho_c$ is a constant parameter that denotes the density at which the bounce takes place in a cosmological context. $\rho_c=2|\sigma|$, where $\sigma$ is the brane tension and the modulus is taken to consider the absolute magnitude as the brane tension is negative in the SS braneworld model. The terms in the modified EFE containing $U$ and $P$ are the effective stress-energy components due to the Weyl contribution, where $U$ and $P$ denote the energy density and pressure due to bulk contribution on the brane, respectively. It is very interesting to note that although we consider a perfect fluid with isotropic pressure as the matter source on the brane, the effective radial and tangential pressures on the brane arising from higher dimensional bulk contributions are different, contributing to an induced pressure anisotropy amounting to $\frac{12P}{\rho_c}$ on the brane, due to bulk effects. Also, we can see that the average effective pressure due to the projected Weyl tensor can be obtained as $\tilde{P}_{eff}^{avg}=\frac{1}{3}\bigg(\tilde{P}^{r}_{eff}+2\tilde{P}^{t}_{eff}\bigg)=\frac{1}{3}\tilde{\rho}_{eff}$, which shows the fact that the additional effective matter appearing in the modified EFE from the projected Weyl term has the EoS of radiation due to the projected Weyl tensor being traceless.
	
	The conservation equation on the (3+1)-brane is the same as in GR as the stress-energy is conserved separately on the brane and bulk.
	So, for the line element given by Eq. (1), we have
	\begin{equation}
		\frac{dp}{dr}=-\frac{1}{2}\frac{d\nu}{dr}(p+\rho).
	\end{equation}

	The line element for a static, spherically symmetric wormhole has the form
\begin{equation}
	ds^2=-e^{\nu(r)}dt^2+\frac{dr^2}{1-\frac{b(r)}{r}}+r^2(d\theta^2+sin^2\theta d\phi^2), \label{eq12}
\end{equation}
where $b(r)$ and $\nu(r)$ denote the shape function and redshift function of the wormhole, respectively.

For the line element Eq. (10), the modified field equations on the 3-brane given by Eqs. (6)-(8) reduce to

\begin{eqnarray}
	\frac{b^\prime}{r^2}=\rho \left(1-\frac {\rho}{\rho_c}\right)-\frac{12 U}{\rho_c}, \label{eq13}\\
	\left(1-\frac{b}{r}\right)\left(\frac{\nu^\prime}{r}+\frac{1}{r^2}\right) -\frac{1}{r^2}
	=p -{\frac {\rho\left(2 p +\rho \right)} {\rho_c}}-\frac{4U}{\rho_c}-\frac{8P}{\rho_c},\label{eq14}\\
	\left(1-\frac{b}{r}\right)\left(\nu'' + {{\nu^\prime}^2} +\frac{\nu^\prime}{r}  \right) -\frac{b^\prime -b}{2r}\left({\nu^\prime} +\frac{1}{r} \right)
	=p-{\frac{\rho \left(2 p +\rho \right)}{\rho_c}}-\frac{4U}{\rho_c}+\frac{4P}{\rho_c}. \label{eq15}
\end{eqnarray}

	The matter on the brane is taken to be a perfect fluid such that the stress energy tensor has the form $T^{\mu}_{\nu}=diag(-\rho,p,p,p)$,
where the pressure and energy density are related via the equation of state (EoS) parameter $\mu$ as
\begin{equation} \label{eq17}
	p(r)=\mu \rho(r).
\end{equation}
The parameter $\mu$ can be evaluated from the Israel Darmois junction conditions at the wormhole surface.

As already discussed, we follow an approach~\cite{S9} where we do not care what happens in the bulk, so we have a lot of freedom for choosing the specific form of the Weyl projection on the brane. Such choices have also been made in~\cite{Sengupta5,Sengupta2}. Here also, we assume a linear EoS connecting $U$ and $P$ of the form $P=\omega U$, where $\omega$ is a constant EoS parameter that we will obtain from the junction conditions. The energy density due to the Weyl projection on the brane is assumed to have a profile $U(r)=\sqrt{\rho_0 \rho(r)}$, where $\rho_0$ is another constant model parameter to be obtained from the boundary conditions at the wormhole surface.

\subsection{Model 1}

The first metric potential which contains the redhift function is assumed to be the Kuchowicz metric function~\cite{Kuchowicz}
\begin{equation}
	e^{\nu(r)}=e^{Br^2+2\ln C}.
\end{equation}
Here $B$ is an arbitrary constant having dimension $[L^{-2}]$ and $C$ denotes a dimensionless constant. The reason behind the choice of
the Kuchowicz potential as the redshift function is that it is a well behaved regular function for all finite radial distances and can
represent the metric potential at the interior of regular collapse solution like the gravastar and other compact objects~\cite{Biswas2020}.
It has also been used as redshift function for wormhole on other instance~\cite{Sengupta2}. As we shall see later, the value of the parameter
$B$ and $C$ as evaluated from the boundary conditions at the wormhole surface ensure the asymptotic flatness of the wormhole such that the
redshift function remains finite for infinitely large radial distances.

\subsubsection{Solution for the Wormhole Shape Function}

Using the EoS in the stress energy conservation equation for the assumed form of the redshift function, the energy density of matter
	constituting the wormhole can be found to be
	\begin{equation} \label{eq18}
		\rho(r)=C_1e^{-\frac{H_1 Br^2}{2\mu}},
	\end{equation}
	where $C_1$ is an integration constant.
	
	\begin{figure*}[thbp] \label{rho}
		\centering
		\includegraphics[width=0.45\textwidth]{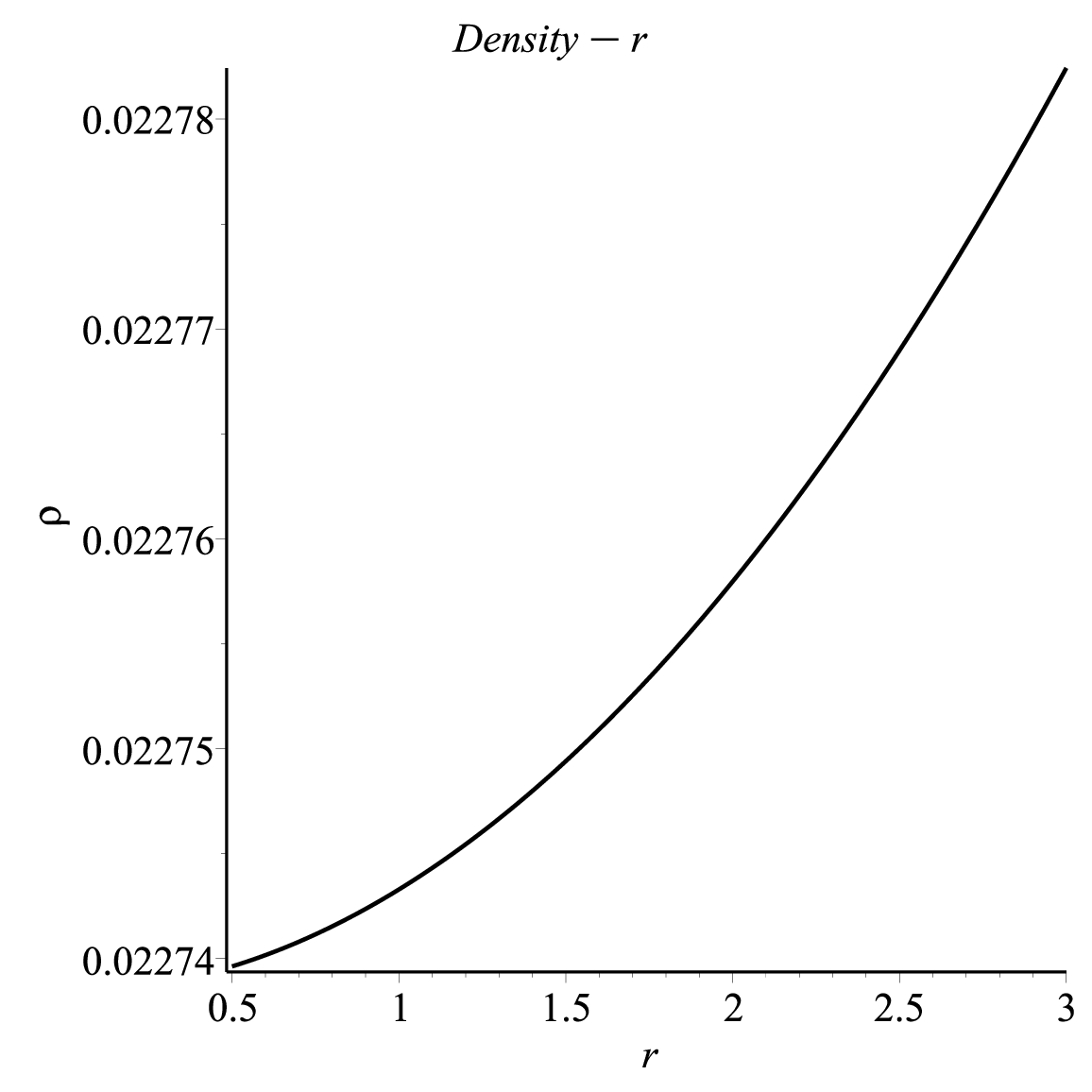}
		\includegraphics[width=0.45\textwidth]{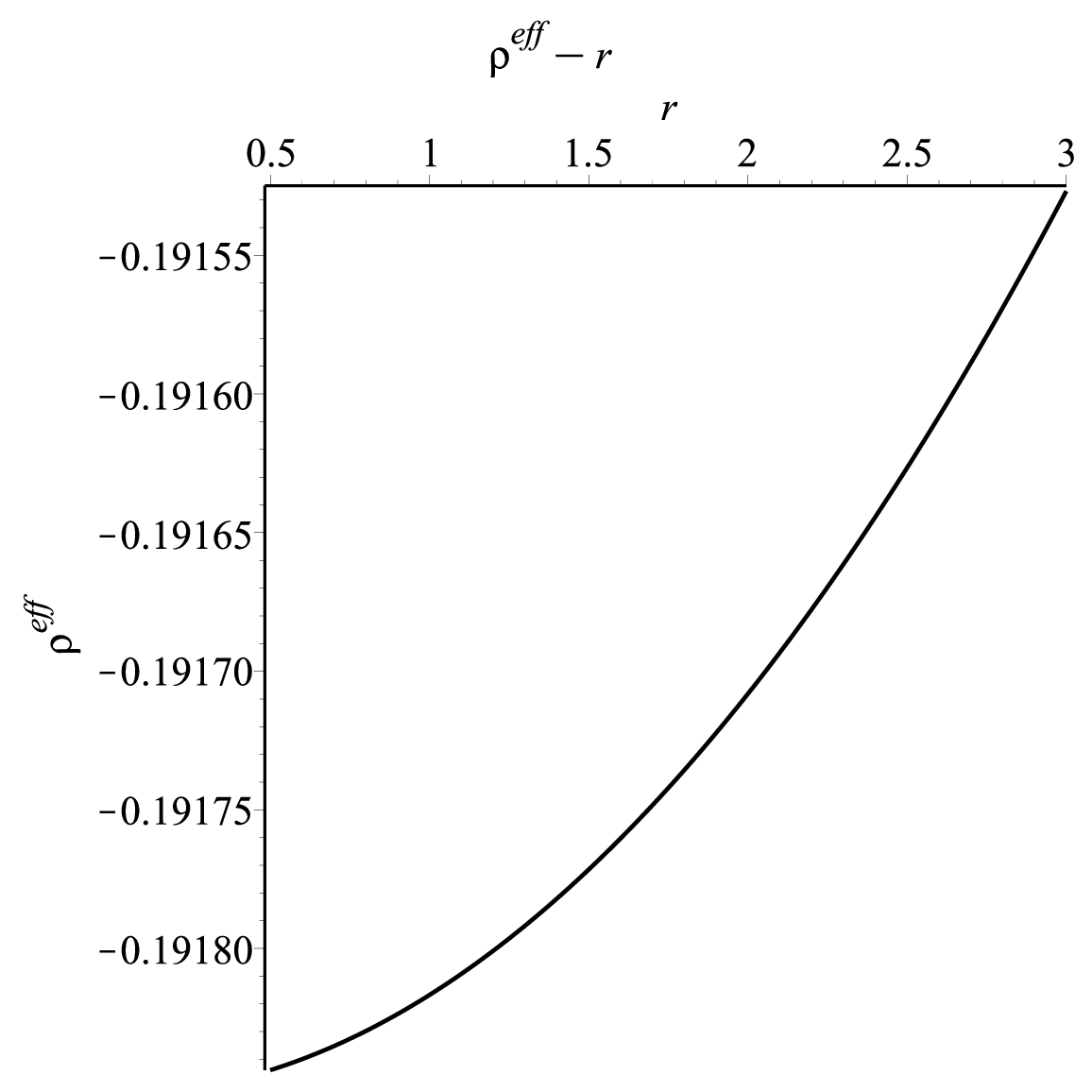}
		\caption{Variation of the energy density and effective energy density with respect to $r$.}
	\end{figure*}
	
	In figure 1, we have plotted the variation of the energy density and the effective energy density along the radial distance. Its significance has been discussed in the concluding section.
	
	Plugging in the obtained energy density in the field Eq. (13) and using the EOS of Eq. (14) along with Eq. (15), we obtain a solution
	for the shape function $b(r)$, having the form
	
	\begin{eqnarray}
		b(r) &=&r{\rm e}^{-2Br^2} \left( \frac{8\pi {{C_1}}^{2} F_2 \mu{{\rm e}^{{\frac {Br^2H }{\mu}}}}}{B \rho_c H G} + \frac{ G {{\rm e}^{2Br^2}} +C_2}{G}  -\frac{112\pi {C_1}{\mu}^{2} {{\rm e}^{{\frac {Br^2 F }{2\mu}}}}}{B  F  F_1 G} -{\frac {16 \left( \omega-1 \right)  \mu \sqrt{\rho_{{0}}{C_1}}{{\rm e}^{{\frac {Br^2 F_1 }{4\mu}}}} }{ B \rho_c   F_1 G}} \right),  \label{eq19}
	\end{eqnarray}
	where $C_2$ is the constant of integration whose value can be obtained from the junction conditions.
	
	Here we have $G=(2Br^2+1),\ F=(3\mu-1),\ H=(\mu-1),\ F_1=(7\mu-1), \ H_1=(\mu+1),  F_2=(2\mu+1)$.
	
	\begin{figure*}[thbp] \label{SF}
		\centering
		\includegraphics[width=0.5\textwidth]{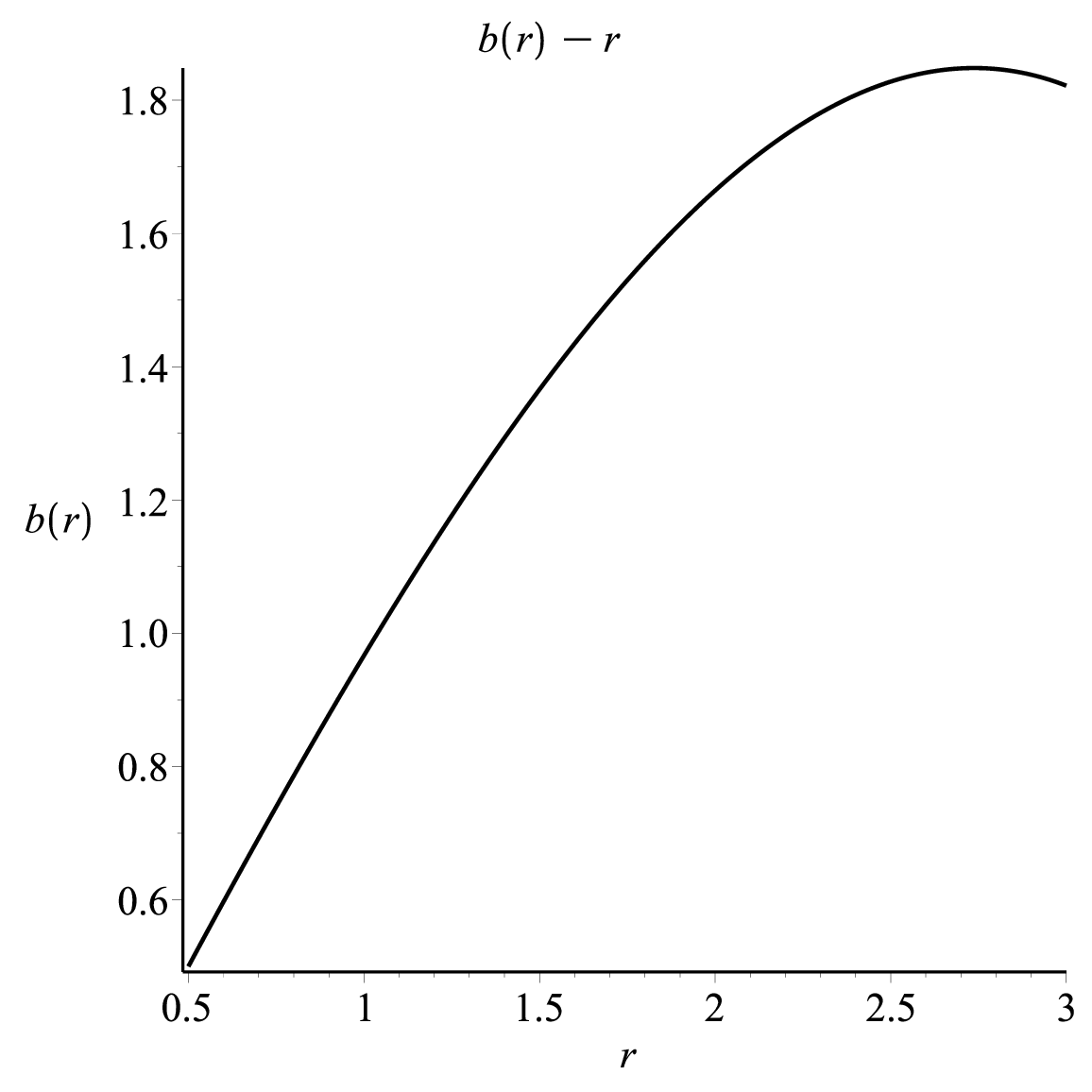}
		\caption{Variation of the shape function with respect to $r$.}
	\end{figure*}
	
	The variation of the shape function along the radial distance is plotted in Fig.2 and as we can see it has a desired shape capable
	of representing a wormhole surface. There is a slight bend as we move away from the throat towards the wormhole surface, whose physical significance is discussed later on. The throat radius is assumed to be $r_0=0.5 km$.
	
	\subsubsection{Validity of NEC}
	
	On the braneworld, there is a modification in the geometry sector of the EFE which can effectively be represented as an effective matter
	sector. So, even for ordinary perfect fluid matter, the components of the stress-energy tensor are different from usual GR. As a result,
	there is an effective $T_{\mu \nu}$ appearing on the RHS of the modified EFE, arising out of the modification to the standard action due
	to the extra dimensional effects. We have $T^{\mu (eff)}_{\nu}=diag(-\rho^{eff}, p_r^{eff}, p_t^{eff}, p_t^{eff})$.
	
	For a braneworld with a timelike extra dimension, the components of the effective stress-energy tensor are given as
	\begin{equation}
		\rho^{eff}=\rho  \left( 1-\frac {\rho }{\rho_c} \right) -\frac{12U}{\rho_c}, \label{eq20}
	\end{equation}
	
	\begin{equation}
		p^{eff}_r=\left(p  +{\frac {\rho  \left( p -\frac{\rho}{2} \right)}{\frac{\rho_c}{2}}}\right)-\frac{4U}{\rho_c}-\frac{8P}{\rho_c}, \label{eq21}
	\end{equation}
	and
	\begin{equation}
		p^{eff}_t=8\pi \Bigg(p-{\frac{\rho \left(p+\frac{\rho}{2}\right)}{\frac{\rho_c}{2}}}\Bigg) -\frac{4U}{\rho_c}+\frac{4P}{\rho_c}.
	\end{equation}
	
	On adding the effective energy density and pressure, we have
	\begin{eqnarray}
		\rho^{eff}+p^{eff}_r={\frac {1}{ \rho_c } \left( -16\pi {{C_1}}^{2} H_1  \left( {{\rm e}^{-{\frac { H_1 Br^2}{2\mu}}}} \right) ^{2}+8\pi {C_1} \rho_c  H_1 {{\rm e}^{-{\frac { H_1 Br^2}{2\mu}}}}+8\sqrt{\rho_{{0}}{C_1}}{{\rm e}^{{\frac{H_1 Br^2}{4\mu}}}} \left( \omega-2 \right)  \right)} \label{eq22}
	\end{eqnarray}
	
	\begin{eqnarray}
		\rho^{eff}+p^{eff}_t={\frac {1}{\rho_c} \left( -16\pi {{C_1}}^{2} H_1  \left( {{\rm e}^{-{\frac { H_1 B{r}^{2}}{2\mu}}}} \right) ^{2}+8\pi {C_1}\rho_c H_1 {{\rm e}^{-{\frac { H_1 B{r}^{2}}{2\mu}}}}+4\sqrt {\rho_0{C_1}}{{\rm e}^{{\frac { H_1 B{r}^{2}}{4\mu}}}} \left( \omega-4 \right)  \right) }
	\end{eqnarray}
	\begin{figure*}[thbp] \label{NEC}
		\centering
		\includegraphics[width=0.45\textwidth]{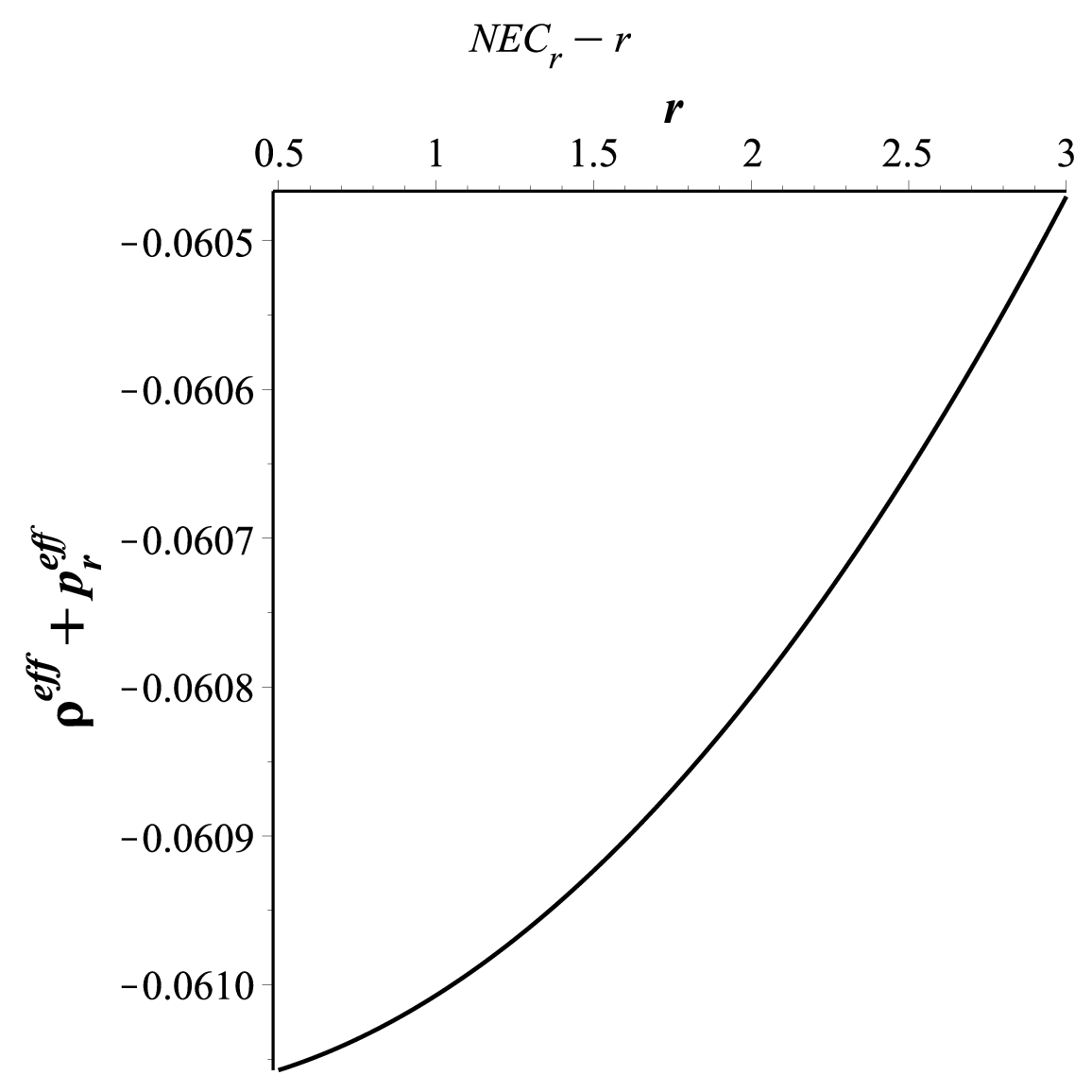}
		\includegraphics[width=0.45\textwidth]{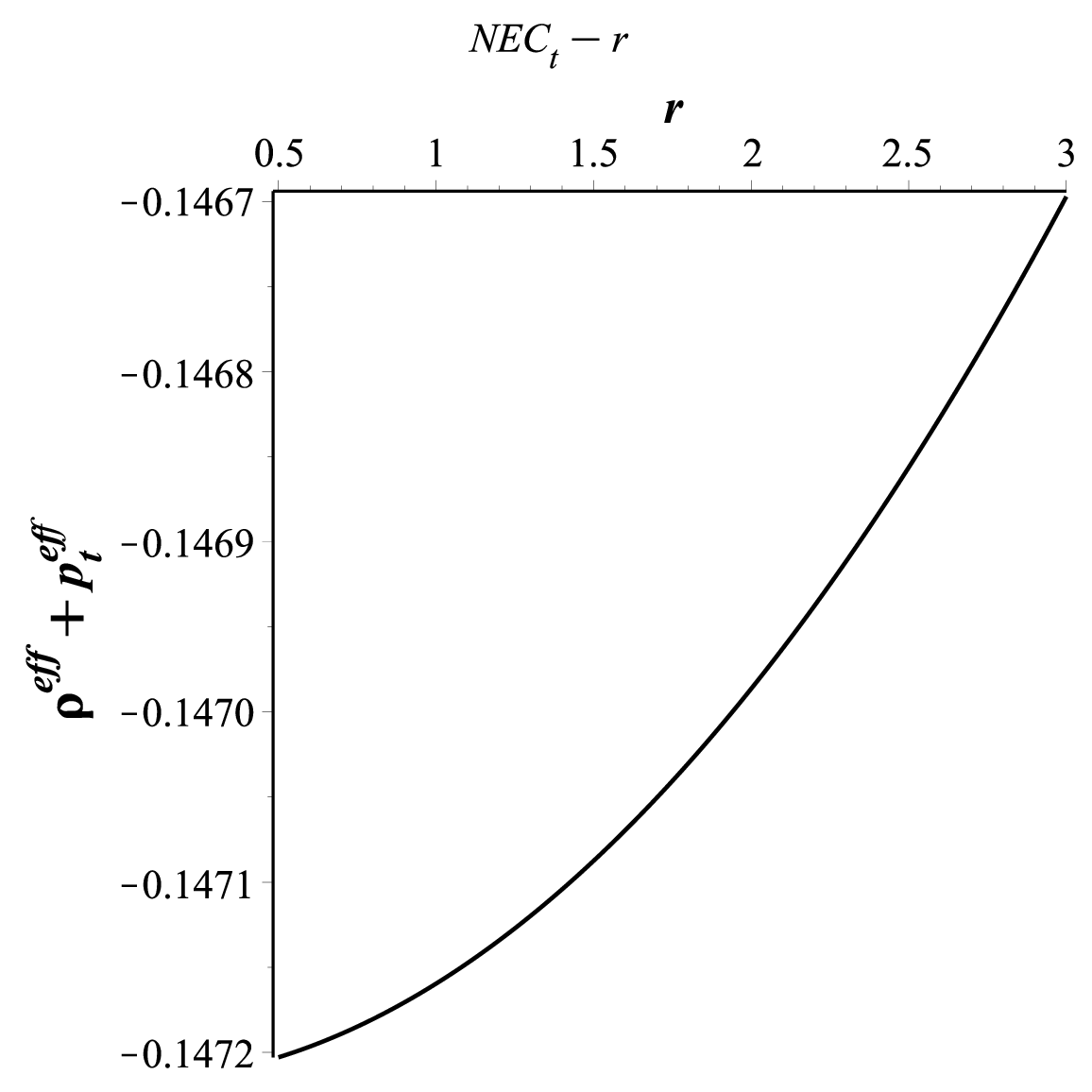}
		\caption{Variation of the NEC with respect to $r$.}
	\end{figure*}
	
	The variation of the sum of effective pressures and density for matter obeying a linear EoS, constituting the wormhole on the braneworld with a
	timelike extra dimension, along the radial distance is plotted in Fig. 3. On using the values of the model parameters obtained from the junction
	conditions, it turns out that the NEC is violated effectively.

	\subsubsection{The Junction Conditions}
	We can see from the variation of the energy density $\rho$, as we approach the surface of the wormhole there is an increase in the matter
	density which indicates the presence of matter at the surface of the wormhole, which however, does not disturb the asymptotic flatness of the wormhole as the flare-out condition is not violated since we are concerned with the effective matter on the brane and not the physical matter. This gives rise to an extrinsic discontinuity at the surface,
	resulting to the generation of intrinsic surface energy density and surface pressure. The surface of the wormhole acting as a junction
	between its interior and exterior spacetimes leads to the geodesic completeness of the wormhole, characterized by a perfect fluid matter
	configuration. The junction conditions suggest a smooth matching of the interior and exterior spacetimes at the junction, involving a
	continuity of the metric potentials, which however does not ensure the continuity of derivatives of the potentials. Thus, the surface
	stress-energy can be obtained following the prescription of Darmois and Israel~\cite{Israel,Darmois}.
	
	The intrinsic surface stress energy tensor $S_{i}^{j}$ is given by the Lanczos equation~\cite{Lanczos1924,Sen1924,Perry1992,Musgrave1996} as
	\begin{equation}\label{eq24}
		S_{j}^{i}=-\frac{1}{8\pi} (\kappa_{j}^{i}-\delta_{j}^{i} \kappa_{k}^{k}),
	\end{equation}
	where discontinuity in the second fundamental form is given by
	\begin{equation}\label{eq25}
		\kappa_{ij}=\kappa_{ij}^{+}-\kappa_{ij}^{-}.
	\end{equation}
	We obtain the second fundamental form using the expression
	\begin{equation}\label{eq26}
		\kappa_{ij}^{\pm}=-n_{\nu}^{\pm}\left[\frac{\partial^{2}X_{\nu}}{\partial \xi^{i}\partial\xi^{j}}+
		\Gamma_{\alpha\beta}^{\nu}\frac{\partial X^{\alpha}}{\partial \xi^{i}}\frac{\partial X^{\beta}}{\partial
			\xi^{j}} \right]|_S,
	\end{equation}
	such that the unit normal vector $n_{\nu}^{\pm}$ have the form
	\begin{equation}\label{eq27}
		n_{\nu}^{\pm}=\pm\left|g^{\alpha\beta}\frac{\partial f}{\partial X^{\alpha}}\frac{\partial f}{\partial X^{\beta}}
		\right|^{-\frac{1}{2}}\frac{\partial f}{\partial X^{\nu}}.
	\end{equation}
	Also, $n^{\nu}n_{\nu}=1$ and $\xi^{i}$ denotes the intrinsic coordinate of the wormhole surface having $f(x^{\alpha}(\xi^{i}))=0$ as its parametric equation. $+$ describes the spacetime exterior to the wormhole, while $-$ describes the interior spacetime of the wormhole.
	
	The solution at the exterior of the wormhole is a vacuum solution but as we have accounted for the projected bulk Weyl tensor on the brane to have a non-zero contribution, so the vacuum will possess a tidal charge which will effectively modify the line element for the vacuum, which is given by
	\begin{equation}
		ds^2=-\left(1-\frac{2M}{r}-\frac{Q}{r^2}\right)dt^2+\left(1-\frac{2M}{r}-\frac{Q}{r^2}\right)^{-1}dr^2+r^2(d\theta^2+\sin^2 \theta d\phi^2),
	\end{equation}
	where $M$ represents the mass of the wormhole and $Q$ represents the tidal charge which is a dimensionless quantity arsing from the effective stress-energy components of the projected Weyl term on the brane. The higher dimensional bulk gravitational effect transferred by the projection of the Weyl tensor to the brane not only modifies the dynamics inside the wormhole but also in the vacuum exterior to its surface boundary. If this correction  to the effective energy density in the exterior spacetime is $U_e$, then $U_e\sim\frac{Q}{r^4}.$ So, the tidal charge parameter can be both greater than or less than zero in accordance with the sign of the effective energy density $U_e$ that arises from the effective extra-matter like contribution due to the projection of the bulk Weyl tensor on the brane. A positive $U_e$ corresponds to a positive $Q$ and vice-versa. Generally for a braneworld with a spacelike extra dimension, the quantity $U_e$ is a negative one to ensure confinement of the gravitational field to the brane and so is $Q$ but as we are interested in a braneworld with a timelike bulk signature, so we had assumed the positive root solution to the effective energy density $U$ inside the wormhole as well. The negative sign before the $Q$ term is also a consequence of the extra dimension being timelike. The parameter $Q$ having a small non-zero positive or negative value introduces significant extra-dimensional UV corrections to standard GR results. In the SS braneworld scenario, the sign of the $U_e$ term will be positive for confining gravity to the brane aided by the negative brane tension and so the tidal charge is also positive. Had it been negative as in the RS scenario, the exterior spacetime would have resembled to a Reisnner-Nordstrom one.
	
	Thus, the surface density is given by
	
	\begin{eqnarray}\label{28}
		\qquad\hspace{-1.0cm}\Sigma  &=&-\frac{1}{2\pi r}\bigg[\sqrt{e^\lambda}\bigg]_-^+=\frac{1}{2\pi r} \left(\sqrt{1-{\frac{2M}{r}}-{\frac{Q}{r^2}}}-\right.\nonumber\\
		\qquad\hspace{-1.40cm}&&\sqrt{1-{\rm e}^{-2Br^2} \left( \frac{8\pi {{C_1}}^{2} F_2 \mu{{\rm e}^{{\frac {Br^2H }{\mu}}}}}{B \rho_c H G} + \frac{ G {{\rm e}^{2Br^2}} +C_2}{G}  -\frac{112\pi {C_1}{\mu}^{2} {{\rm e}^{{\frac {Br^2 F }{2\mu}}}}}{B  F  F_1 G} -{\frac {16 \left( \omega-1 \right)  \mu \sqrt{\rho_{{0}}{C_1}}{{\rm e}^{{\frac {Br^2 F_1 }{4\mu}}}} }{ B \rho_c   F_1 G}} \right)} ,
	\end{eqnarray}

	The surface pressure has the form
	\begin{eqnarray}
		\qquad\hspace{-2.0cm}&&\mathcal{P}  =\frac{1}{16\pi r } \bigg[\bigg(\frac{2f+f^\prime r}{\sqrt{f}}\bigg) \bigg]_-^+ =\frac {3{{\rm e}^{-2Br^2}}}{ G ^{2} F  \rho_c H {r}^{4}B\pi } \left( B \left( \frac{F G^2}{12} \right)   \rho_c H  \left( 2Mr^2-{r}^{3}+ \left( Q-M \right) r-Q \right){{\rm e}^{Br^2}}\right.\nonumber\\
		&&\left.\sqrt{{ \left( \frac{2{\mu}^{2}{C_1} {{\rm e}^{\frac{3Br^2}{2} }} {{\rm e}^{{\frac {Br^2}{2\mu}}}}}{ B F  G }- \frac{ C_2  {{\rm e}^{{\frac {Br^2}{\mu}}}}}{ G }+\frac{F_2 {{\rm e}^{Br^2}}\mu{{C_1}}^{2}}{ B \rho_c  H  G }\right)   {{\rm e}^{{-\frac{Br^2}{\mu}}}} }} +\sqrt{1-{\frac{2M}{r}}-{\frac{Q}{r^2}}}\left( -\frac{{{\rm e}^{{\frac {Br^2 F }{2\mu}}}}\mu \rho_c {C_1}H }{6} \left( {r}^{3} H_1 {B}^{2}-G\mu\right.\right.\right.\nonumber\\
		&&\left.\left.\left. +\frac{B ( 5\mu+1) r}{2} \right)   \frac{F}{3}  \left(  \left( {r}^{3} H_1 {B}^{2}-\frac{G \mu}{2}+\frac{rB(3\mu+1)}{2} \right)  \frac{{C_1}^{2} F_2 {{\rm e}^{{\frac {Br^2H }{\mu}}}}}{2}+B \rho_c H {C_2} \left( {B}^{2}{r}^{3}-\frac{G}{4}+Br \right)  \right)  \right) {r}^{3} \right) \nonumber\\
		&&\frac {{{\rm e}^{Br^2}}}{\sqrt {{\frac {1}{B \rho_c  FH  G } \left( 2{\mu}^{2}{C_1} \rho_c {{\rm e}^{\frac{3 Br^2}{2}}}H {{\rm e}^{{\frac {Br^2}{2\mu}}}}- \left( B{C_2} \rho_c H {{\rm e}^{{\frac {Br^2}{\mu}}}}+\frac{2F_2  \mu{{C_1}}^{2} {{\rm e}^{Br^2}}}{2} \right)  \left( \frac{F}{3} \right)  \right)  \left( {{\rm e}^{{-\frac{Br^2}{\mu}}}} \right)}}}
		{\frac {1}{\sqrt{1-{\frac{2M}{r}}-{\frac{Q}{r^2}}}}}.
	\end{eqnarray}

	A static wormhole is characterized by a vanishing surface energy density and surface pressure $\Sigma =\mathcal{P}=0$ at the boundary, yielding the condition
	\begin{equation}\label{eq30}
		b(r)|_{r=R}=2M+\frac{Q}{R}.
	\end{equation}
	This is one of the boundary conditions that we shall use to obtain the unknown constant model parameters. In addition, the junction conditions imply the metric potential $g_{tt}$ and its derivative $\frac{\delta g_{tt}}{\delta r}$ to be continuous across the surface boundary at $r=R$.
	
	On choosing physically realistic values of the model parameters $ M=0.5789993753 M_\odot $,  $\  Q = 0.008$, $\  \rho_c = 0.41m^4$, $r_0=0.5km$ and $R=3km$ we obtain $B = -0.000125km^{-2}$, $\  \mu = 0.41$, $\rho_0 = 0.14$, $ C1=0.02273840355$, $ C2= -5681.061699$ and $\ \omega = 0.3450530290$. The significance of the vital physical parameters shall be discussed in the concluding section.
	
	\subsubsection{Tidal acceleration}
	
	It is essential to constrain the velocity of the traveller at the throat of the wormhole by setting up a realistic limit on the tidal forces at the throat, as infinitely large tidal forces would lead to a Weyl curvature singularity at the throat resulting in a pinch-off, that would rip the traveller apart. We take a realistic upper limit on the tangential and radial components of the tidal acceleration to be the acceleration due to gravity on the Earth.
	
	The radial component of the tidal acceleration can be computed in terms of the Riemann curvature tensor as
	\begin{equation}
		|R_{rtrt}|=|(1-\frac{b}{r})\big[\frac{\nu"}{2}+\frac{\nu'^2}{4}-\frac{b'r-b}{2r(r-b)}.\frac{\nu'}{2}\big]|\leq g_{Earth}.
	\end{equation}
	
	The tangential component of tidal acceleration using the Riemann tensor is given by
	\begin{equation}
		\gamma^2 |R_{\theta t \theta t}|+\gamma^2v^2 |R_{\theta r \theta r}|=|\frac{\gamma^2}{2r^2}\big[v^2(b'-\frac{b}{r})+(r-b)\nu'\big]|\leq g_{Earth}.
	\end{equation}
	
	Here, the Lorentz factor is given as $\gamma=\frac{1}{\sqrt{1-v^2}}$ and $v$ denotes the  velocity of the traveller traversing the throat of the wormhole. On a realistic note, the traveller being a macroscopic object, velocity of the traveller must be much less than unity. So, $\gamma\approx 1$ appears to be quite a reasonable approximation. Using the assumed redshift function in the form of the Kuchowicz potential and the shape function obtained by solving the modified field equations on the brane, an upper limit on the velocity of the traveller traversing the wormhole throat can be obtained using the above inequality as
	\begin{equation}\label{99}
		v\leq 0.05341107\sqrt{g_{Earth}}.
	\end{equation}
	We also have a reasonably small radial tidal acceleration at the wormhole throat.
	
	\subsubsection{Linearized stability analysis}
	
	In this section we shall perform a qualitative linearized stability analysis, for which we shall consider the throat radius of the wormhole to depend on the proper time. We consider the throat radius $r_0=x(\tau)$. On such a consideration, the energy density $\Sigma$ and pressure $\mathcal{P}$ can be expressed by the equations
		\begin{equation}
		\Sigma=-\frac{1}{2\pi x}\sqrt{f(x)+\dot{x}^2},
	\end{equation}
	and
	\begin{equation}
		\mathcal{P}=\frac{1}{8\pi}\frac{f'(x)}{\sqrt{f(x)}}-\frac{\Sigma}{2},
	\end{equation}
	
	such that $f(x)=1-\frac{2M}{x}-\frac{Q}{x^2}$ with the parameter $M$ denoting the wormhole mass and $Q$ the tidal charge.
	
	The conservation of energy-momentum yields the equation of motion
	\begin{equation}
		\dot{x}^2+V(x)=0.
	\end{equation}
	
	Here, the potential $V(x)$ can be expressed as
	\begin{equation}\label{LS2}
		V(x)=f(x)-[2\pi x \Sigma (x)]^2.
	\end{equation}
	
	In order to perform a stability analysis, we need to consider a linearization around an assumed static solution $x_0$ for the equation of motion given by Eqn. (36).
	\begin{figure*}[thbp] \label{NEC}
		\centering
		\includegraphics[width=0.5\textwidth]{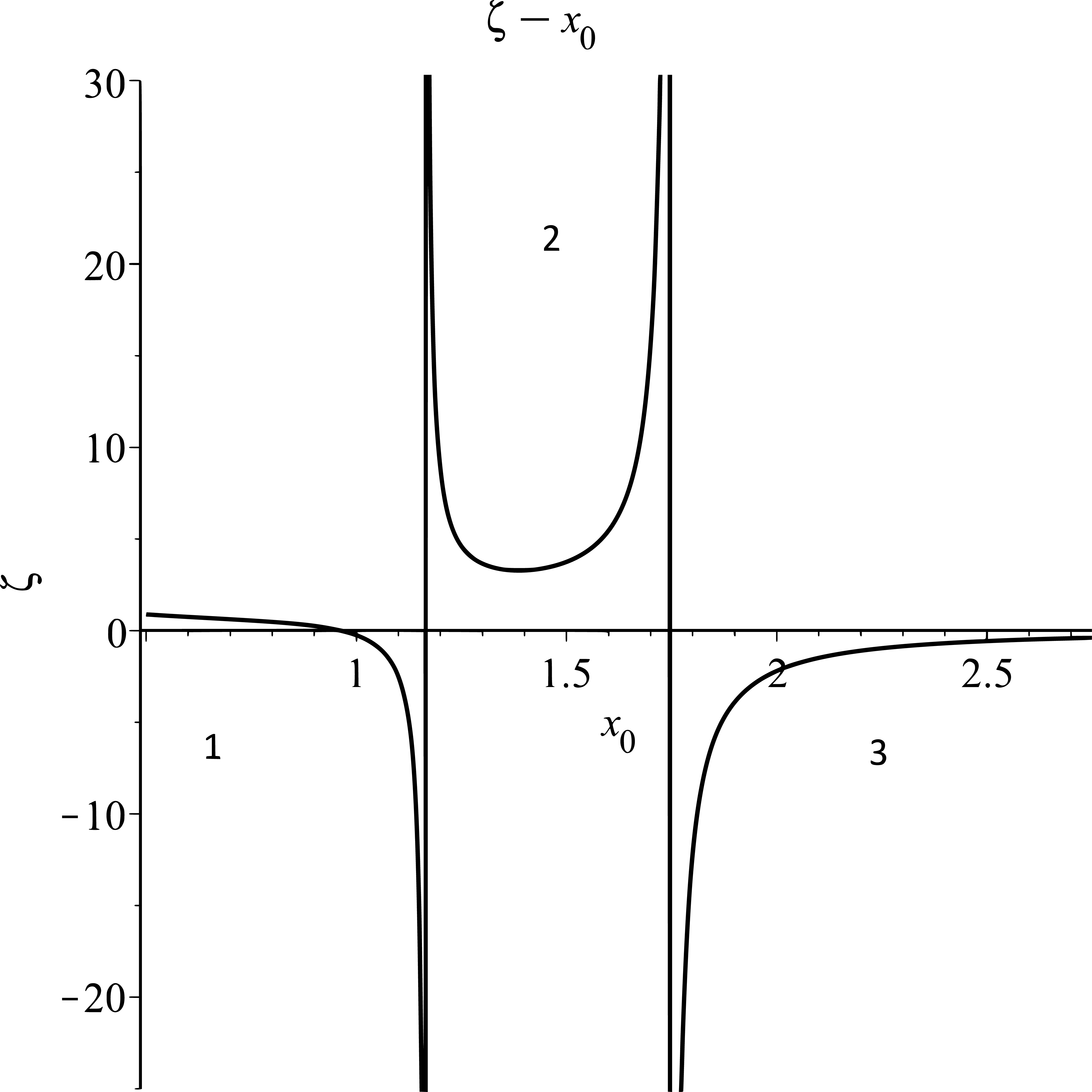}
		\caption{Plot of $\zeta$ vs $x_0$.}
	\end{figure*}
	
	On Taylor expansion of the potential $V(x)$ upto the second order around $x_0$, we have
	\begin{equation}\label{LS3}
		V(x)=V(x_0)-V'(x_0)(x-x_0)+\frac{1}{2}V"(x_0)(x-x_0)^2+O[(x-x_0)^3],
	\end{equation}
	prime representing derivative with respect to x.
	
	As we consider a static wormhole on the brane, described by a time independent line element, so both the potential and its first derivative with respect to x shall vanish at the assumed static solution for the equation of motion $x_0$. Thus the wormhole will be stable only for $V"(x_0) > 0$. We introduce a parameter $\zeta=\frac{\delta \mathcal{P}}{\delta \Sigma}$, in terms of which we obtain the stability condition for the wormhole.
	
	The second derivative of the potential with respect to $x$ can be expressed in terms of the energy density, pressure, parameter $\zeta$ and the function $f(x)$ as
		\begin{equation}\label{LS5}
		V''(x)=f''(x)-8 \pi^2 [ (\Sigma +2\mathcal{P})^2+ \Sigma (\Sigma+\mathcal{P})(1+2\zeta)
	\end{equation}
	
	Thus, the stability condition for the wormhole in terms of the parameter $\beta$ is
	\begin{equation}\label{LS7}
		\zeta< \frac{\frac{f''(x_0)}{8\pi^2}-(\Sigma +2\mathcal{P})^2-2\Sigma(\Sigma+\mathcal{P})}{4 \Sigma (\Sigma +\mathcal{P})}.
	\end{equation}
	
	Plugging in the expressions for $\Sigma$ and $\mathcal{P}$, the above inequality reduces to
	
	\begin{equation}\label{LS8}
		\zeta< \frac{x_0^2 (f_0')^2-2x_0^2 f_0'' f_0}{4 f_0(x_0 f_0' -2f_0)}-\frac{1}{2}
	\end{equation}
	
	The parameter $\beta$ turns out to have the form
	\begin{eqnarray}
		\zeta={\frac{-2{r}^{4}+ \left(-12\pi +9 \right) M{r}^{3}+ \left( \left(-16\pi +5 \right) Q+20{M}^{2} \left(\pi -\frac{1}{2} \right) \right) r^2+36 \left( \pi -{\frac{11}{36}} \right) M Q r+12{Q}^{2} \left( \pi -\frac{1}{4} \right) }{8r \left( 2Mr-r^2+Q \right) \pi  \left(3Mr-r^2+2Q \right) }}
	\end{eqnarray}
	
	The stable regions for our wormhole model have been indicated as regions 1,2 and 3 in Fig. 4 whose significance is discussed briefly in the concluding section.
	
	\subsubsection{Surface Redshift}
	
	As we see our wormhole model can be constituted with normal matter that obeys the null and strong energy conditions and hence the presence of a photon cannot be ruled out. If any such photon is emitted from the surface of the wormhole, it will experience a redshift as it travels from a lower to higher gravitational potential due to loss in energy in escaping the gravitational field of the wormhole. This is a consequence of the fact that photons have a non-zero gravitational mass.
	
	The surface redshift of the wormhole can be be obtained by using the formula
	\begin{eqnarray}\label{eq35}
		Z_{s}&=&-1+\frac{1}{\sqrt {g_{\it tt}}} =-1+{\frac {1}{\sqrt {{C}^{2}{{\rm e}^{Br^2}}}}}.
	\end{eqnarray}
	
	\begin{figure*}[thbp]
		\centering
		\includegraphics[width=0.5\textwidth]{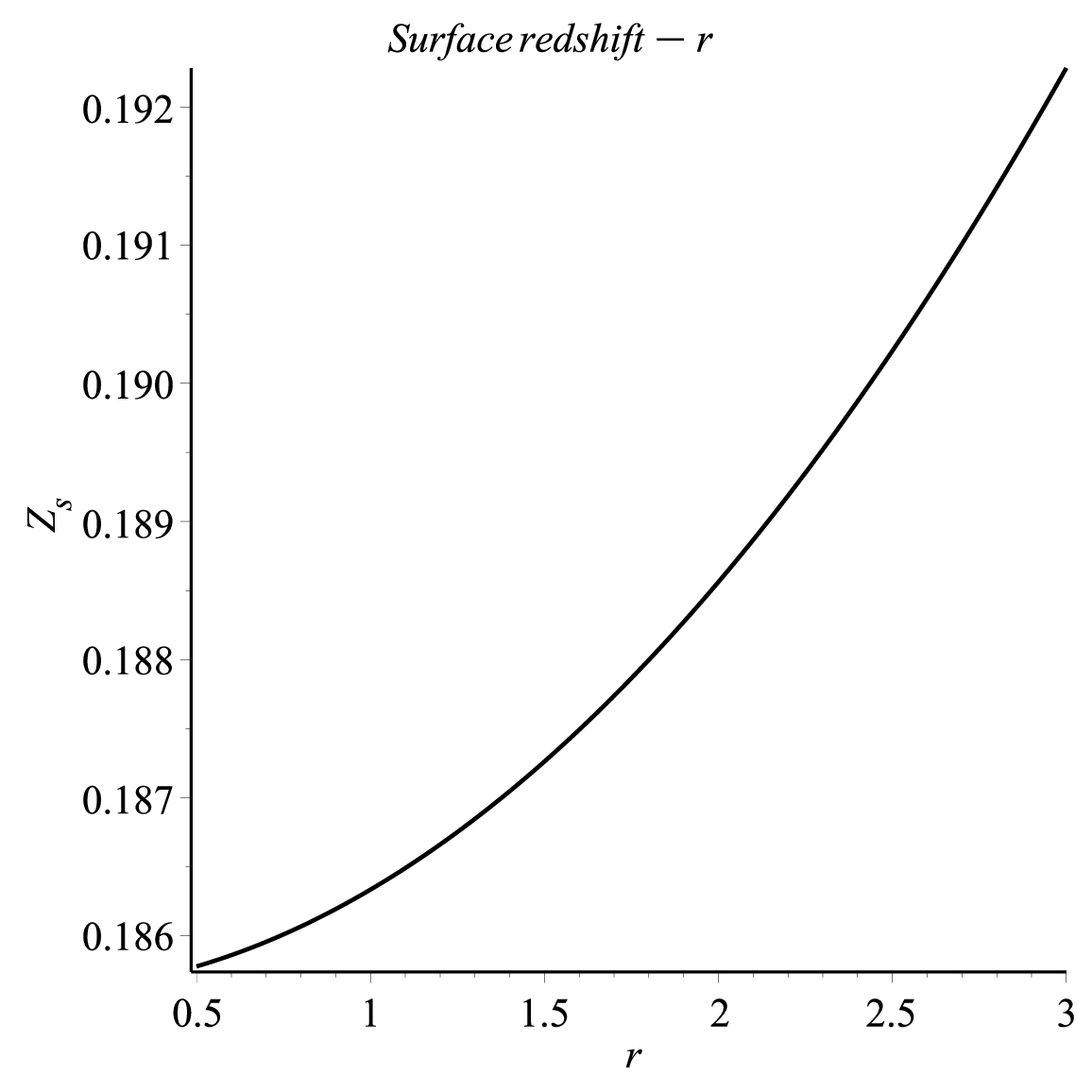}
		\caption{Variation of the Surface redshift of the wormhole with respect to $r$.}
	\end{figure*}
		The surface redshift has been plotted along the radial distance in Fig. 5. For a vanishing cosmological constant, the surface redshift $(Z_s)$ must not exceed the value 2 in order to ensure stability~\cite{Bohmer2006}. The value however should not exceed 5 in case of the presence of a $\Lambda$-term. As we can see from Figure 5, the surface redshift of the wormhole obtained indicates towards its stability.
	
	\subsubsection{Acceleration and nature of the wormhole}
	
	The 'attractive' or 'repulsive' geometrical nature of a wormhole can be inferred upon using the redshift and shape functions function, involving derivative of the former with respect to the radial coordinate. By 'attractive' wormhole geometry, it means that the radial component of the four-acceleration of a static observer is positive, implying the observer requires an outward-directed radial acceleration to refrain from being pulled into the wormhole. On the contrary, a 'repulsive' wormhole geometry is characterized by a negative radial four-acceleration, which physically means the requirement of an inward-directed radial acceleration on the static observer to prevent being pushed away from the wormhole.
	
	The four-velocity of a static observer in terms of the redshift function can be written as
	\begin{equation}
		U^{\mu}=\frac{dx^{\mu}}{d\tau}=(e^{-\frac{\nu(r)}{2}},0,0,0),
	\end{equation}
	$\tau$ representing the proper time. In turn, the four-acceleration $a^{\mu}=U^{\mu}_{;\nu}U^{\nu}$ has its radial component given by
	\begin{equation}
		a^r=\frac{\nu'}{2}\bigg(1-\frac{b(r)}{r}\bigg).
	\end{equation}
	For a test particle initially at rest, the geodesic equation in the radial direction is
	\begin{equation}
\frac{d^2 r}{dt^{\tau}}=- \Gamma^r_{tt}\bigg(\frac{dt}{d\tau}\bigg)^2= -a^r.
	\end{equation}
	The radial four acceleration for our wormhole model can be computed to be
	\begin{eqnarray}	
		a^r &=&  Br\left( -{\rm e}^{-2Br^2} \left( \frac{8\pi {{C_1}}^{2} F_2 \mu{{\rm e}^{{\frac {Br^2H }{\mu}}}}}{B \rho_c H G} + \frac{C_2}{G}  -\frac{112\pi {C_1}{\mu}^{2} {{\rm e}^{{\frac {Br^2 F }{2\mu}}}}}{B  F  F_1 G} -{\frac {16 \left( \omega-1 \right)  \mu \sqrt{\rho_{{0}}{C_1}}{{\rm e}^{{\frac {Br^2 F_1 }{4\mu}}}} }{ B \rho_c   F_1 G}} \right)\right){v}^{2}
	\end{eqnarray}
	\begin{figure*}[thbp] \label{Radacc}
		\centering
		\includegraphics[width=0.5\textwidth]{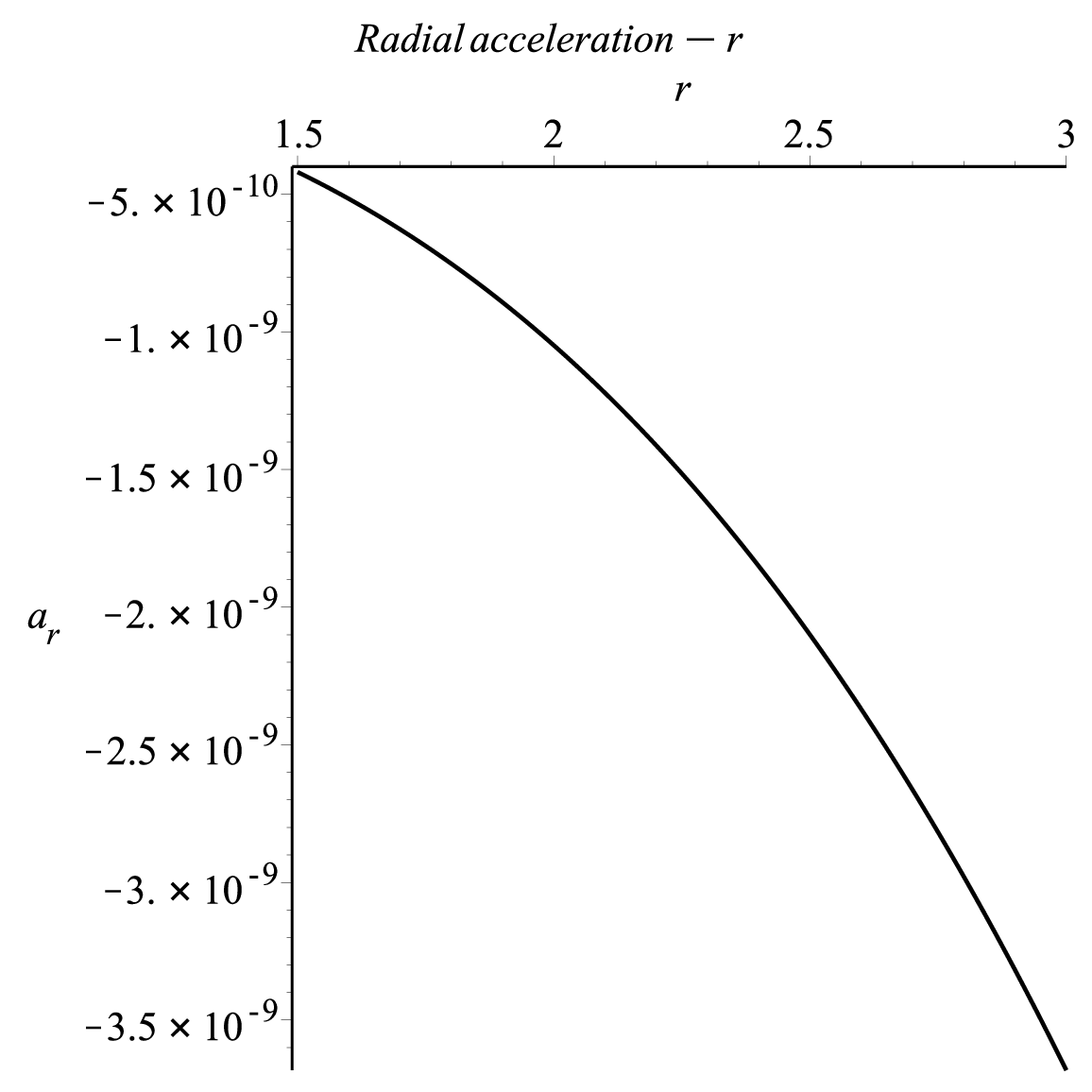}
		\caption{Plot of radial component of acceleration with respect to $r$.}
	\end{figure*}
	A variation of the radial four-acceleration along the radial distance has been plotted in Fig. 5. We find that the radial four-acceleration turns out to be negative which suggests that our wormhole spacetime has a repulsive geometric nature as opposed to wormhole on a brane with spacelike extra dimension.
	
\pagebreak
\subsection{Model 2}

Since we have complete freedom of choice for the matter on the brane, the wormhole solution obtained by us in model 1 is not an unique solution. So, we extend our analysis to attempt the construction of another traversable wormhole for a different choice of the redshift function. We choose a form of the redshift function used previously by Anchordoqui et al.~\cite{Anch}. The redshift function has the form $e^\nu(r)=e^{-\frac{2\beta}{r}}$, where $\beta>0$. Here, the parameter $\beta$ is a constant whose value can be determined from the Israel-Darmois junction conditions.

\subsubsection{Solution for the Wormhole Shape Function}

To obtain the energy density of matter constituting the wormhole, we use the form of the EoS for the brane matter in the conservation equation on the brane. Solving the differential equation yields the energy density having the form
\begin{equation}
		\rho(r)=C_1e^{\frac{(\mu+1)\beta}{r\mu}}.
\end{equation}
Here, $C_1$ is the constant of integration.

\begin{figure*}[thbp] \label{rho}
	\centering
	\includegraphics[width=0.45\textwidth]{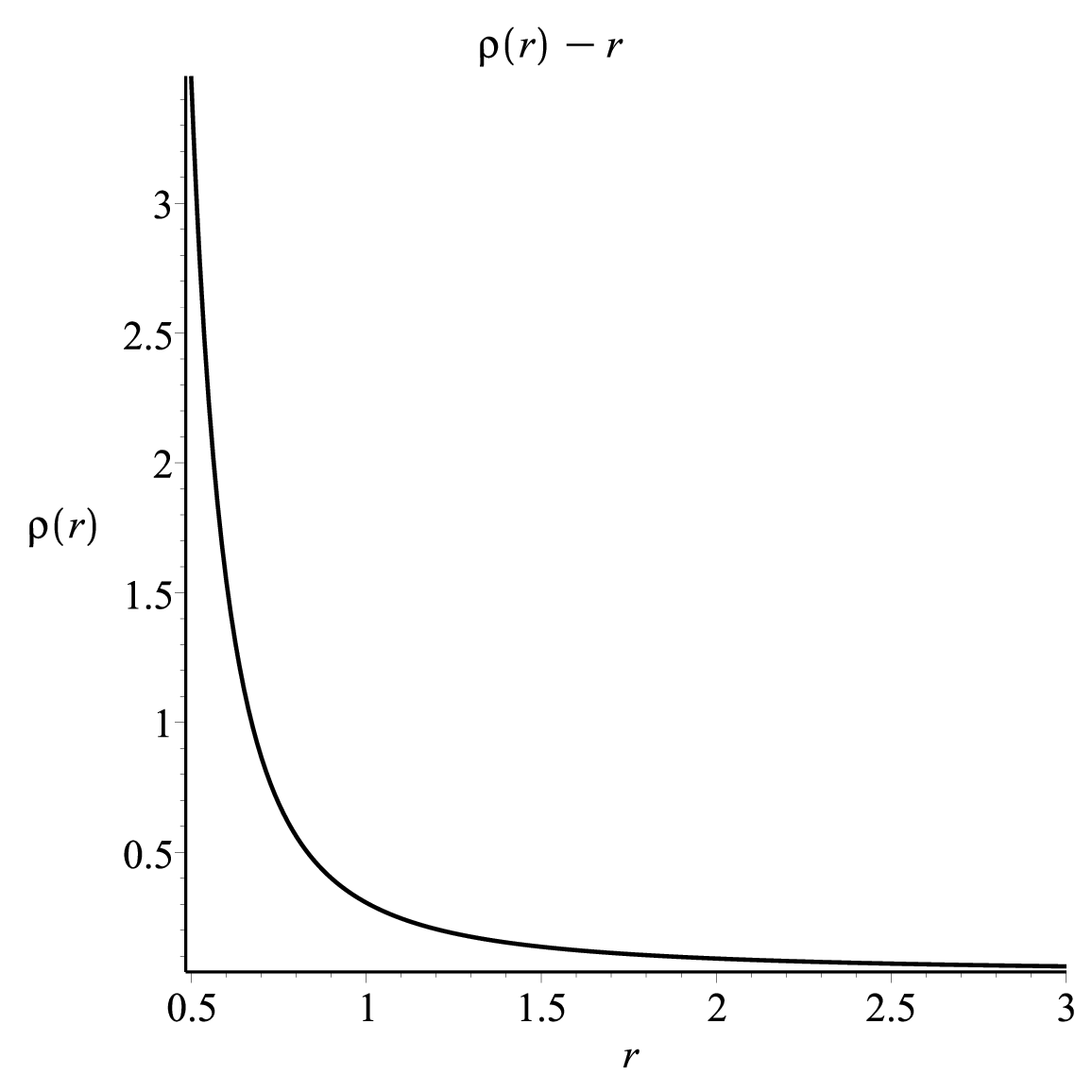}
	\includegraphics[width=0.45\textwidth]{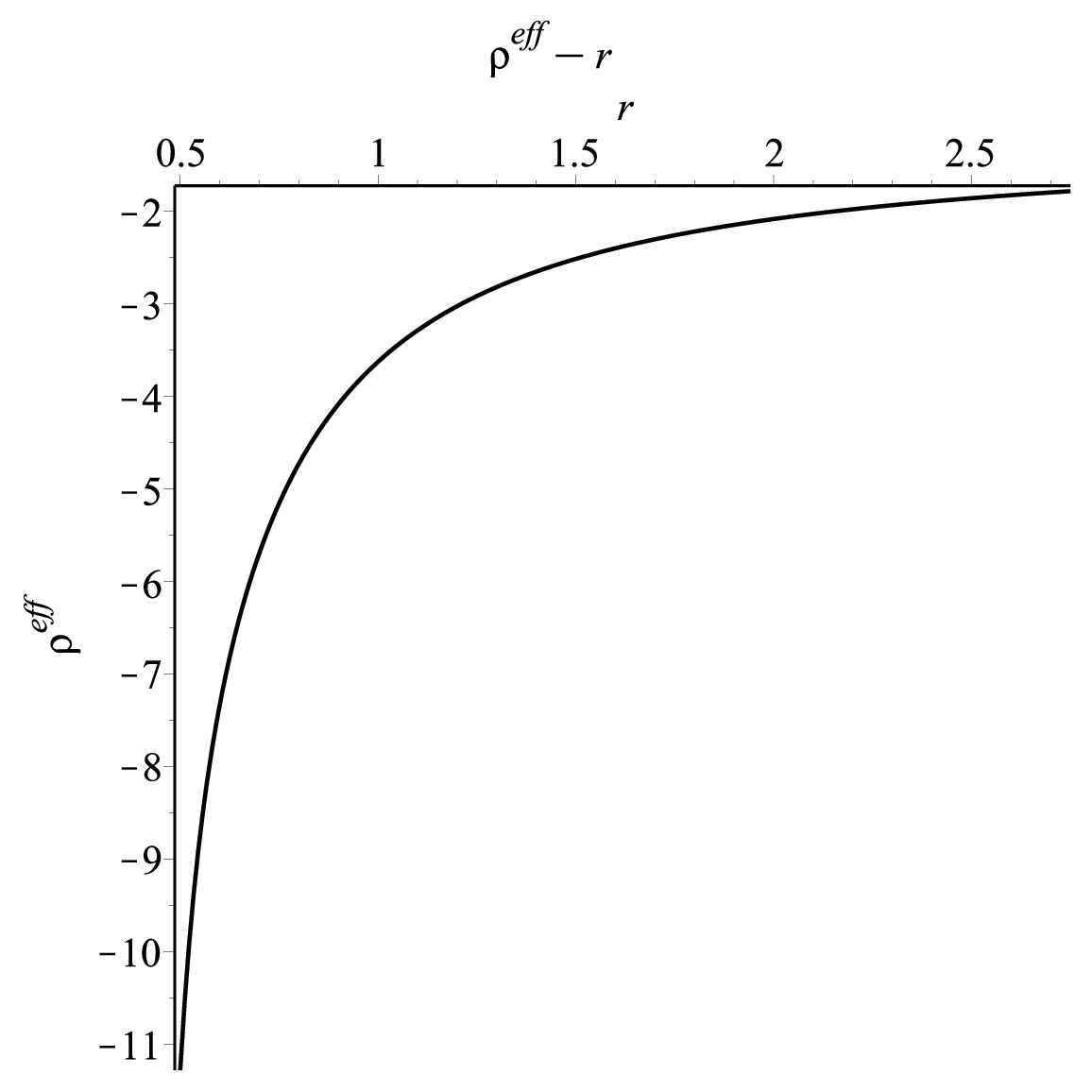}
	\caption{Variation of the energy density and effective energy density with respect to $r$.}
\end{figure*}

In Figure 7 we have plotted the variation of the energy density and effective energy density of the constituent matter along the radial distance from the wormhole throat. The density decreases with radial expanse but this does not gurantee the asymptotic flatness or flare out of the wormhole as, on the brane the energy density is modified due to the effective matter description arising from both the local quadratic correction and the the non-local bulk effect transmitted via the projection of the bulk Weyl tensor behaving effectively as additional matter. So,it has to be ensured that the effective energy density is minimum at the surface of the wormhole. The effective energy density is found to be maximum at the throat and reduces as we move away from the throat, as desired.

Using our chosen form of the redshift function along with the EoS of the brane matter and effective additional matter due to bulk contributions in the modified EFE on the brane, the shape function of the wormhole is computed to have a form

\begin{eqnarray}\label{V1}
	b \left( r \right) ={\frac {8r}{\rho_{{c}} \left( 2\beta+r \right) } \left( \left( \omega+\frac{1}{2} \right)r^3 \sqrt{\rho_{{0}}C_1}{{\rm e}^{{\frac{ (\mu+1)\beta}{2\mu r}}}}+\pi {C_1}^{2} r^3  {\rho_{0}}^{2} \left( 2\mu+1 \right)   {{\rm e}^{{\frac{2(\mu+1)\beta}{\mu r}}}}-\pi {{\rm e}^{{\frac {  (\mu+1) \beta}{\mu r}}}}C_1\mu r^3  \rho_{0}\rho_{{c}}+\frac{\rho_{{c}}}{4}\beta \right) }
\end{eqnarray}

\begin{figure*}[thbp] \label{SF}
	\centering
	\includegraphics[width=0.5\textwidth]{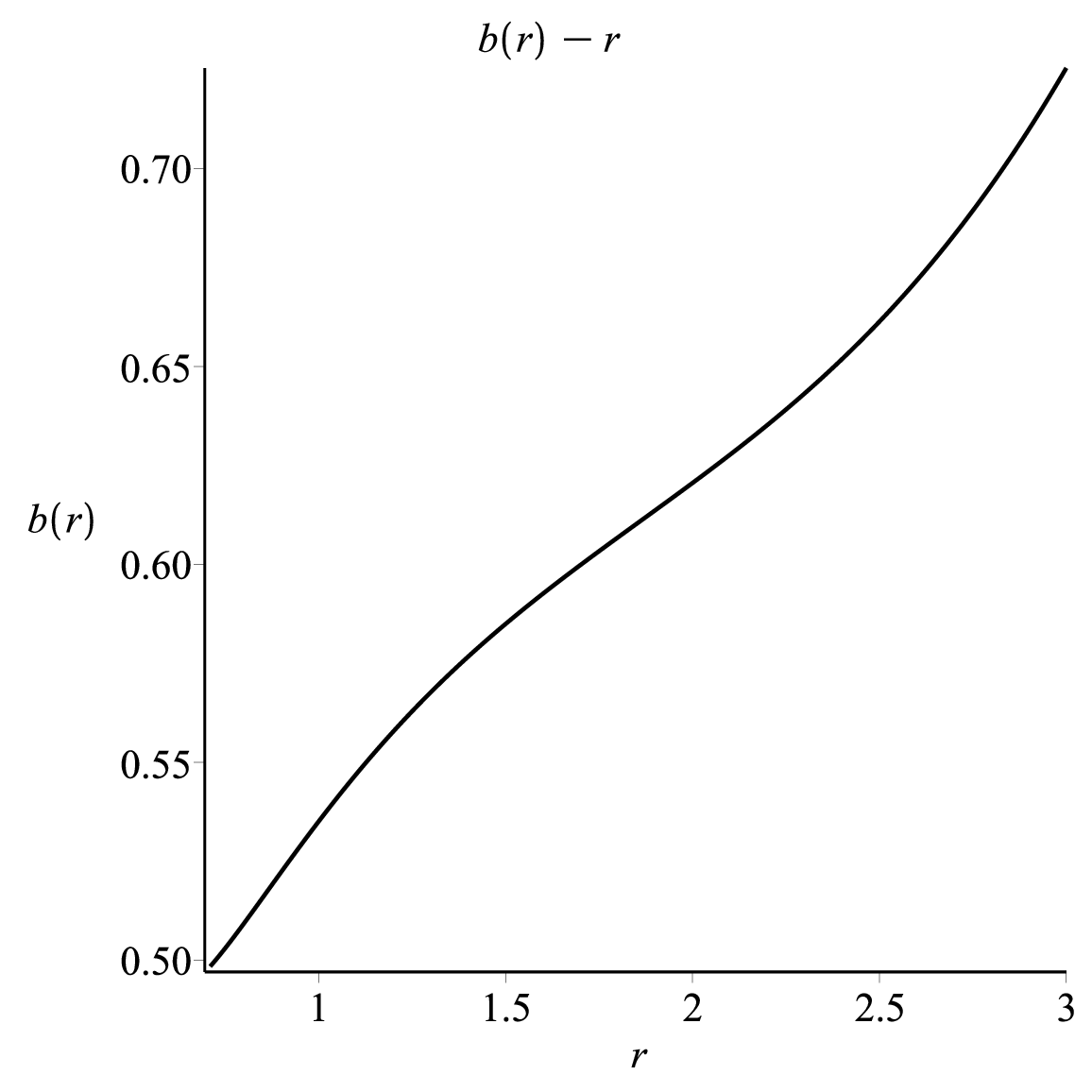}
	\caption{Variation of the shape function with respect to $r$.}
\end{figure*}

The variation of the shape function along the radial extent of the wormhole has been plotted in Figure 8. We find that all the necessary properties for a desired shape function can be satisfied by our obtained function. At the throat, the shape function is equal to the radius of the throat and for all radial distances larger than the throat radius, the ratio of $\frac{b(r)}{r}$ is less than unity. The shape of the wormhole is not exactly identical to model 1 as expected, since we have chosen a different form for the wormhole redshift function. It has a fair resemblance to the shape of a Kuchowicz wormhole on the RS braneworld with a spacelike extra dimension.

\subsubsection{Validity of NEC}

\begin{figure*}[thbp] \label{NEC}
	\centering
	\includegraphics[width=0.45\textwidth]{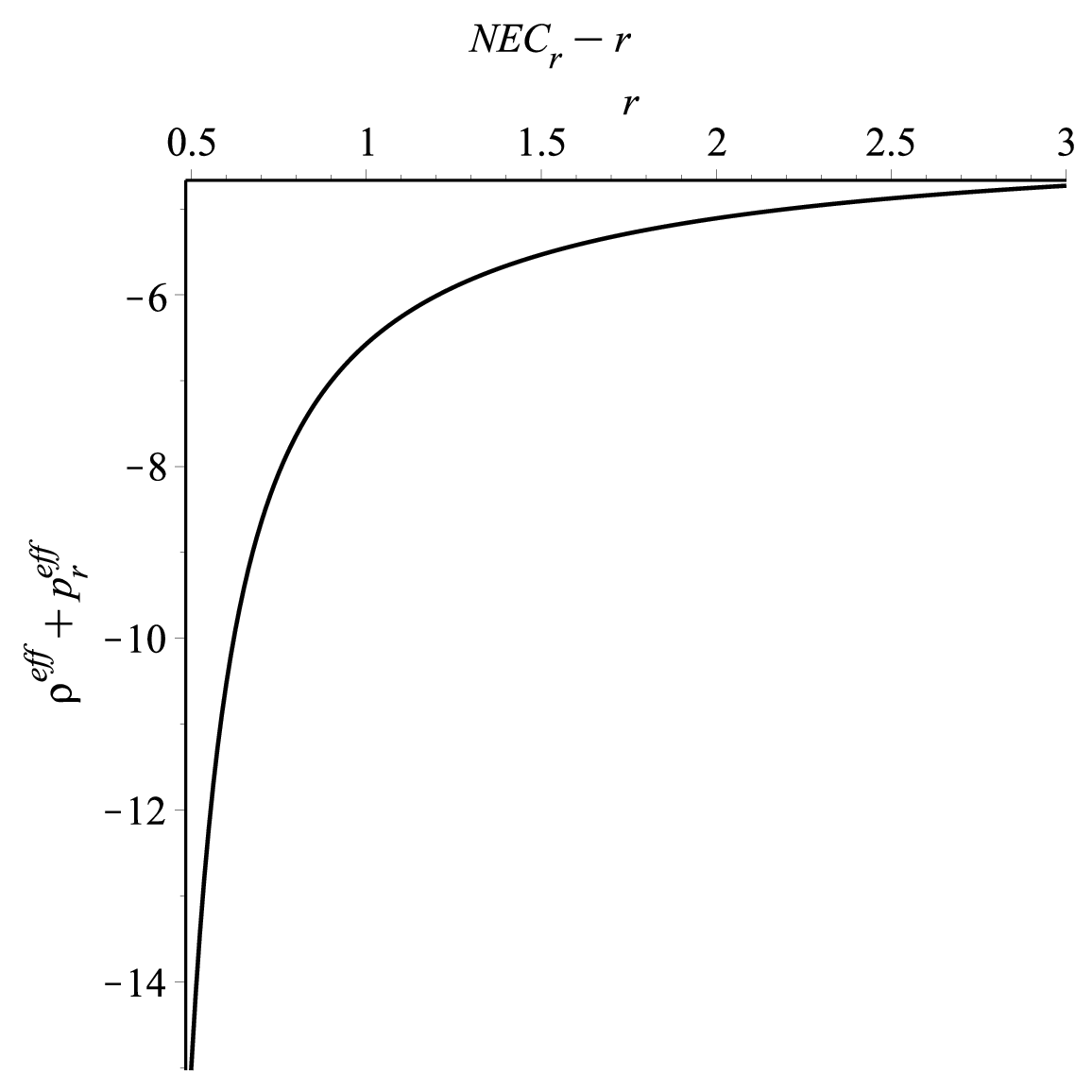}
	\includegraphics[width=0.45\textwidth]{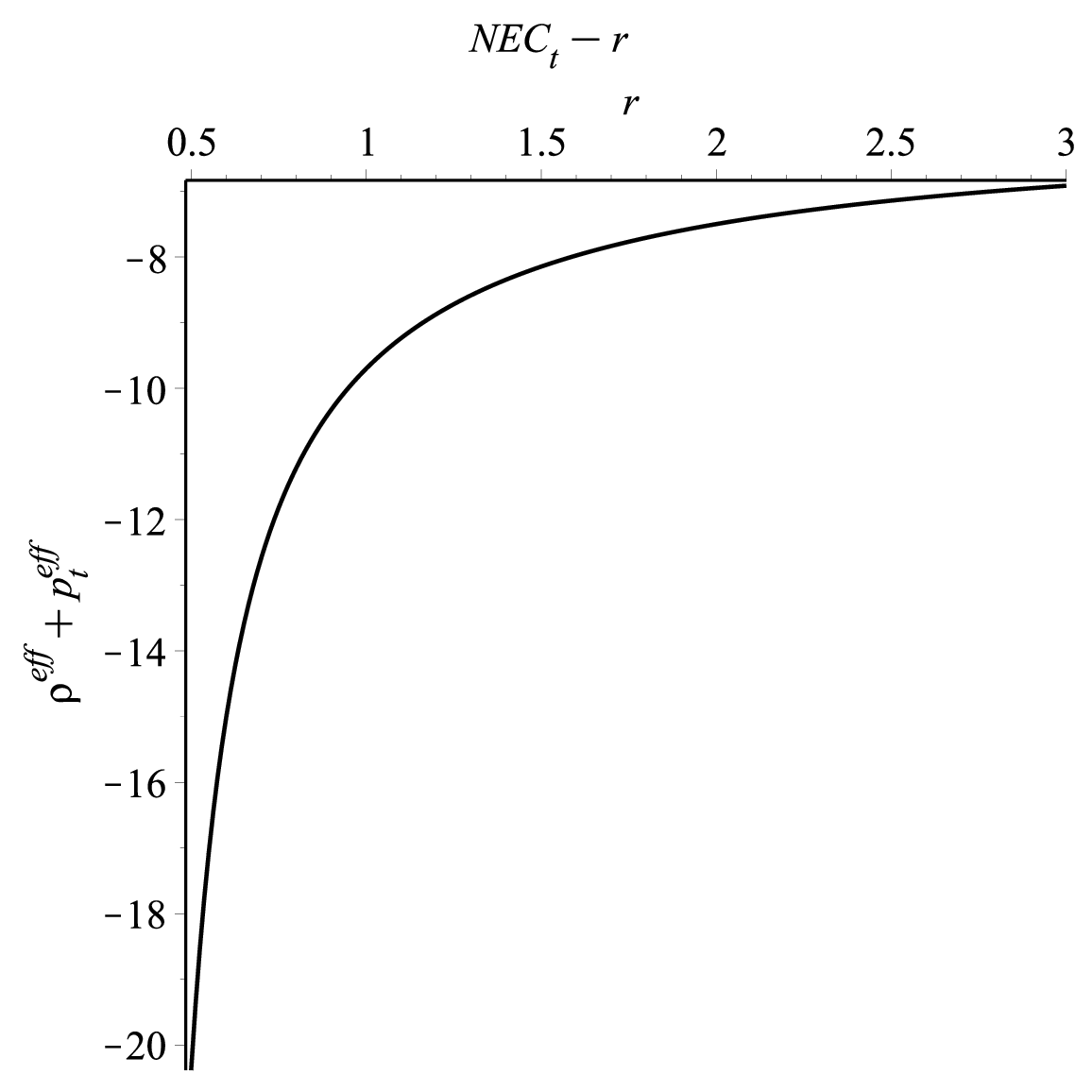}
	\caption{Variation of the NEC with respect to $r$.}
\end{figure*}

As we have considered normal matter on the brane, so the energy conditions will apparently be respected by the matter. However, the local and non-local corrections on the brane in the high energy UV limit will result in an effective matter description on the brane which is expressed in terms of the effective stress-energy tensor as discussed in Model 1 for the Kuchowicz redshift function. Thus, $T^{\mu (eff)}_{\nu}=diag(-\rho^{eff}, p_r^{eff}, p_t^{eff}, p_t^{eff})$.

The effective energy density along with the effective pressures in the radial and tangential directions can be computed from the correction terms introduced in Eqs. (18)-(20). The effective energy density plotted in the right panel of Eq. (7) can be expressed as

\begin{eqnarray}\label{v3}
\rho^{eff}={\frac {1}{\rho_{{c}}} \left( -12\sqrt {\rho_{0}C_1}{{\rm e}^{{\frac { (\mu+1)   \beta}{2\mu r}}}}-8\pi C_1\rho_{0}{{\rm e}^{{\frac{(\mu+1)\beta}{\mu r}}}} \left( \rho_{0}C_1{{\rm e}^{{\frac{\left(\mu+1 \right) \beta}{\mu r}}}}-\rho_{{c}} \right)  \right) }.
\end{eqnarray}

The sum of the effective energy density and effective pressure in the radial direction is computed to have the form

\begin{eqnarray}\label{v2}
\rho^{eff}+p_r^{eff}={\frac {1}{\rho_{{c}}} \left( -8\sqrt {\rho_{0}C_1} \left(\omega+2 \right) {{\rm e}^{\frac{1}{2}{\frac { (\mu+1)   \beta}{\mu r}}}}-16C_1 \left( \rho_{0}C_1{{\rm e}^{{\frac { (\mu+1)  \beta}{\mu r}}}}-\frac{1}{2}\rho_{{c}} \right) (\mu+1)   \rho_{0}{{\rm e}^{{\frac { (\mu+1)   \beta}{\mu r}}}}\pi  \right) }.
\end{eqnarray}

The sum of the effective energy density and effective pressure in the transverse directions can be expressed as

\begin{eqnarray}\label{v4}
\rho^{eff}+p_t^{eff}={\frac {1}{\rho_{{c}}} \left( 4\sqrt {\rho_{0}C_1} \left(\omega-4 \right) {{\rm e}^{{\frac{(\mu+1)\beta}{2\mu r}}}}-16C_1 \left( \rho_{0}C_1{{\rm e}^{{\frac {(\mu+1)\beta}{\mu r}}}}-\frac{1}{2}\rho_{{c}}\right)(\mu+1)\rho_{0}{{\rm e}^{{\frac{(\mu+1)\beta}{\mu r}}}}\pi  \right) }.
\end{eqnarray}

The variation of the sums of effective energy density and effective pressure along the radial expanse of the wormhole have been plotted in Figure 9. The NEC is found to be violated by the effective matter in both the radial and transverse directions. So, the NEC can be violated at the throat by normal matter on the brane having a positive EoS parameter. This is a unique feature of a wormhole on the brane arising due to the corrections to the EFE from the extra dimensional effects.

\subsubsection{The Junction conditions}

As already discussed for our first wormhole model, presence of matter on the surface of the wormhole allows it to act as a junction between the interior and exterior spacetimes leading to geodesic completeness of the wormhole. The extrinsic discontinuity resulting from the matter at the surface is responsible for generating a surface energy density and surface pressure. These two quantities can be computed making use of the Darmois-Israel junction conditions~\cite{Darmois,Israel} as we had done for the previous wormhole model. Despite having a vacuum exterior, the spacetime metric for the exterior is not Schwarzschild but rather of the RN type as discussed earlier. To enforce confinement of bulk gravity close to the brane with a timelike extra dimension characterized by a non-Lorentzian bulk signature, the sign of the tidal charge parameter must be positive as contrasted to the RS brane.

The surface energy density is computed to have the form
\begin{eqnarray}\label{v7}
	\Sigma&=&{\frac {1}{2\pi r} \left( \sqrt {1-{\frac {2m}{r}}-{\frac {Q}{ r^2  }}}\right.}-\nonumber\\
&&{\left.\sqrt {-{\frac {16r}{\rho_{{c}} \left( 2\beta+r \right) } \left( \frac{1}{2}\sqrt {\rho_{0}C_1} r^2  \left(\omega+\frac{1}{2} \right) {{\rm e}^{{\frac { (\mu+1)\beta}{2\mu r}}}}+{C_1}^{2} r^2  {\rho_{0}}^{2}\pi  \left( \mu+\frac{1}{2} \right) {{\rm e}^{{\frac { 2(\mu+1)\beta}{\mu r}}}} -\frac{1}{2}\pi {{\rm e}^{{\frac { (\mu+1)   \beta}{\mu r}}}}C_1\mu r^2  \rho_{0}\rho_{{c}}-\frac{1}{16}\rho_{{c}} \right) }} \right) }.
\end{eqnarray}

The surface pressure turns out to have the form
\begin{eqnarray}\label{v8}
	\mathcal{P}&=&-{\frac {\sqrt {2}}{16 r^4 \left( 2\beta+r \right) ^{2}\rho_{{c}}\pi \mu} \left(\left( 2m r^2  - r^3  + \left( Q-m \right)r-Q \right) \rho_{{c}}\mu \left( 2\beta+r \right) ^{2}\sqrt {8}\right.}\nonumber\\
&&{\left.\sqrt {{\frac {r}{\rho_{{c}} \left( 2\beta+r \right) } \left(\rho_{{c}} \left( \pi{{\rm e}^{{\frac { (\mu+1)\beta}{\mu r}}}}C_1\mu r^2  \rho_{0}+\frac{1}{8} \right)-\left( \omega+\frac{1}{2} \right)  r^2  \sqrt {\rho_{0}C_1}{{\rm e}^{{\frac { (\mu+1)  \beta}{2\mu r}}}}-\left( 2\mu+1 \right) \pi {C_1}^{2} r^2  {\rho_{0}}^{2}{{\rm e}^{{\frac {2 (\mu+1)   \beta}{\mu r}}}}  \right) }}\right.}\nonumber\\
&&{\left.+8 \left( - \left(\mu r^3  +\mu \left( 2\beta+1 \right) r^2  +\frac{\beta}{4} \left( \mu-\frac{1}{11} \right) r-\frac{{\beta}^{2}}{2} (\mu+1)   \right) r \left( \omega+\frac{1}{2} \right) \sqrt {\rho_{0}C_1}{{\rm e}^{{\frac { (\mu+1)   \beta}{2\mu r}}}}-2{{C_1}}^{2}{\rho_{0}}^{2} \left( \mu r^3  +\right.\right.\right.}\nonumber\\
&&{\left.\left.\left.\mu \left( 2\beta+1 \right)  r^2  +\beta r \left( 2 \mu-\right)-{\beta}^{2}(\mu+1)\right) \pi r \left( 2\mu+1 \right) {{\rm e}^{{\frac { 2(\mu+1)   \beta}{\mu r}}}}+\rho_{{c}} \left( C_1\rho_{0} \left( \mu r^3  +\mu \left( 2\beta+1 \right)  r^2  \right.\right.\right.\right.}\nonumber\\
&&{\left.\left.\left.\left.+\frac{ \left( 5\mu-1\right)}{2}  \beta r-{\beta}^{2} (\mu+1)   \right) \pi r{{\rm e}^{{\frac{ (\mu+1)\beta}{\mu r}}}}+\frac{\beta}{8}+\frac{r\beta}{4}+\frac{r^2}{8}\right) \mu \right) r^3 \sqrt{\left(1-\frac{2m}{r}-\frac{Q}{r^2}\right)}  \right)\frac{1}{\sqrt{\left(1-\frac{2m}{r}-\frac{Q}{r^2}\right) }}}\nonumber\\
&&{{\frac{1}{\left(\sqrt{{\frac {r}{\rho_{{c}} \left( 2\beta+r \right) } \rho_{{c}} \left( \pi {{\rm e}^{{\frac{(\mu+1) \beta}{\mu r}}}}C_1\mu r^2  \rho_{0}+\frac{1}{8} \right) - \left( \omega+\frac{1}{2} \right)  r^2  \sqrt{\rho_{0}C_1}{{\rm e}^{{\frac { (\mu+1) \beta}{2\mu r}}}}-\left( 2\mu+1 \right) \pi {C_1}^{2} r^2  {\rho_{0}}^{2}{{\rm e}^{2{\frac {(\mu+1)\beta}{\mu r}}}}  }}\right)}}}.
\end{eqnarray}	

Since we are considering a static wormhole, the surface energy density and pressure must vanish at the boundary which provides one of the boundary conditions for the wormhole. The continuity of the metric along with the radial derivative across the wormhole surface also follows from the junction conditions. We again choose realistic values for the model parameters $  \mu = 0.691,\  R = 3.0km,\  Q = 0.0050953846, r_0=0.496km, \rho_c = 0.41m^4$ to compute the unknown model parameters $M = 0.7066592851M_\odot,\  C_1 = 0.02681286656, \  \omega = -0.4353636831,  \beta = 0.994742337, \rho_0 = 0.064$. Since we are not considering any form of exotic matter with negative pressure, so the EoS parameter for the matter constituting the wormhole cannot be negative. Moreover, as already mentioned, the tidal charge parameter must also be positive to ensure confinement around the brane. This justifies our choice of the model parameters. It is worth mentioning that a small tidal charge arising from the bulk effect significantly modifies the high energy behaviour on the brane. The obtained mass of the wormhole is slightly higher than the mass of the one constructed from the Kuchowicz metric function. The parameter $\beta$ is bounded to be greater than zero and it turns out so from our analysis as desired. The same can be said about the sign of the parameter $\rho_0$. However, the EoS parameter $\omega$ turns out to be negative in this case as contrasted to the Kuchowicz wormhole. This does not cause any problem as this EoS does not represent any real physical matter but an additional effective matter like presence on the brane arising from the projection of the bulk Weyl tensor on the brane. This behaves like some effective exotic matter adding to the composition of the wormhole, but there is no presence of any physical exotic matter that has been considered.

\subsubsection{Tidal acceleration}

We have discuused already that the tidal force on the observer trying to traverse the wormhole should be small enough to ensure smooth passage. It is usually considered that both the radial and tangential components of the tidal acceleration on the observer is not more than the acceleration due to gravity on the earth. Once, we know both the metric potentials of the wormhole, it is possible to compute the tidal accelerations using the components of the Riemann tensor as shown in Eqns (31)-(32). We make a reasonable approximation that the velocity with which the traveller traverses the wormhole is negligeble compared to the velocity of light, which simplifies the Lorentz factor to unity. It turns out that for the obtained values of model parameters, the radial tidal acceleration is less than the acceleration due to gravity on earth and for the tangential acceleration to obey the same condition, the velocity of the traveller is found to be constrained by the inequality
\begin{equation}
	v\leq 0.0997 \sqrt{g_{earth}}.
\end{equation}
We see that the traveller can traverse the wormhole with a greater velocity than for the Kuchowicz wormhole on the brane.

\subsubsection{Linearized stability analysis}

Following our approach of linearized stability analysis for the Kuchowicz wormhole, we compute the parameter $\zeta$ which turns out to have the form

\begin{figure*}[thbp] \label{NEC}
	\centering
	\includegraphics[width=0.5\textwidth]{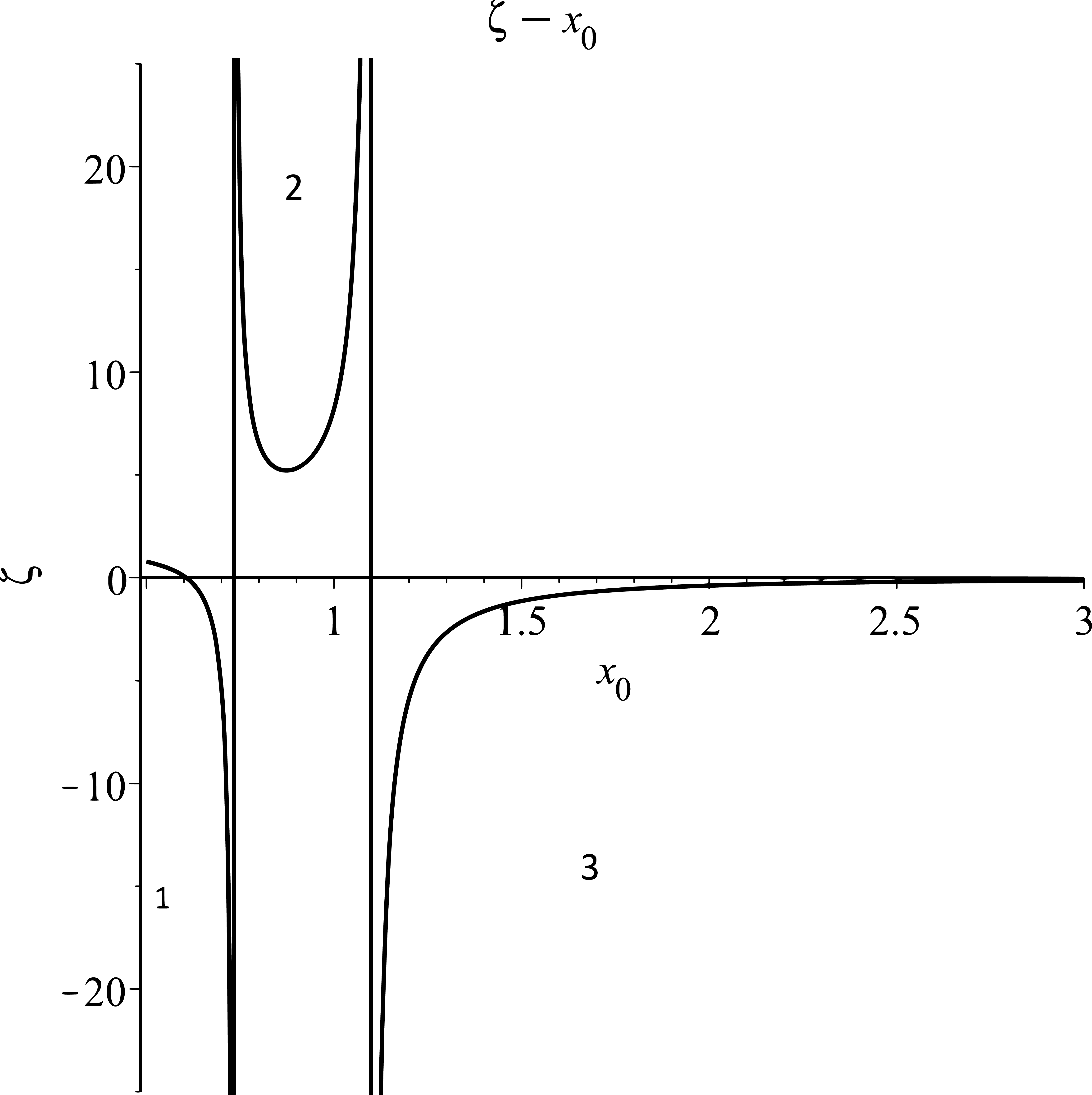}
	\caption{Plot of $\zeta$ vs $x_0$.}
\end{figure*}

\begin{eqnarray}\label{v6}
\zeta=\frac{1}{8}{\frac {-2{r}^{4}+ \left( -12\pi +9 \right) m r^3  + \left( \left( -16\pi +5 \right) Q+20{m}^{2} \left( \pi -\frac{1}{2} \right) \right)  r^2  +36mQ \left( \pi -{\frac {11}{36}} \right) r+12{Q}^{2} \left( \pi -\frac{1}{4} \right) }{r \left( 2mr- r^2  +Q \right) \pi \left( 3mr- r^2  +2Q \right) }}
\end{eqnarray}

Making use of the stability condition (41) in terms of the parameter $\zeta$, we have plotted the variation of $\zeta$ with $x_0$, the assumed static solution of the equation of motion obtained from the energy-momentum conservation, around which the linearization is performed. The regions of stability are marked in the figure as 1, 2, and 3 and represent the regions where the effective potential constructed from the surface density of the wormhole have a minima.

\subsubsection{Surface redshift}

The surface redshift provides another check on the stability of the wormhole. Since, we do not require exotic matter to constitute the wormhole as a result of the brane and bulk effects, a possibly present (produced) photon experiences redshift on escaping the gravitaional field of the wormhole due to loss in energy owing to its non zero gravitational mass, which can computed from the wormhole redshift function as

\begin{figure*}[thbp]
	\centering
	\includegraphics[width=0.5\textwidth]{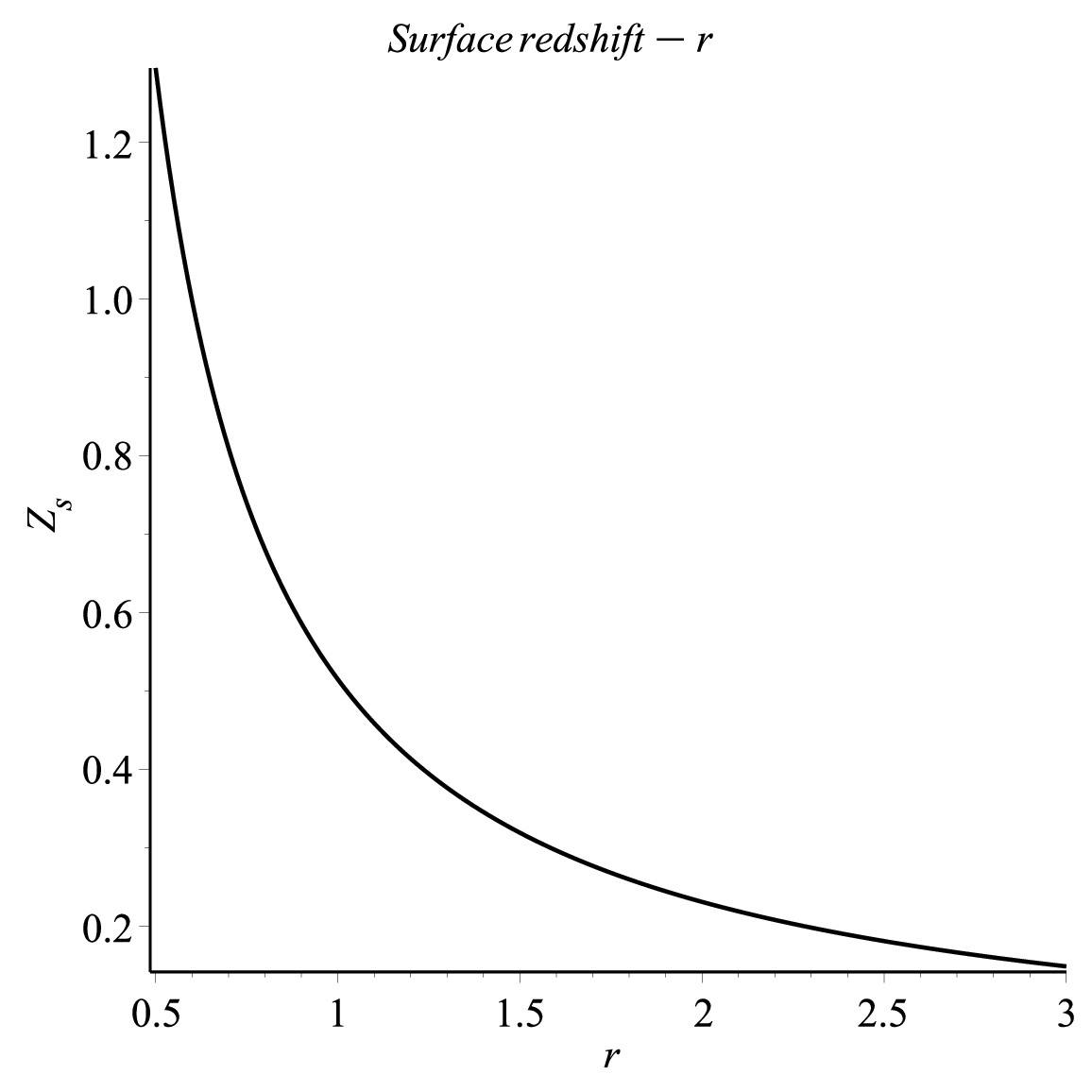}
	\caption{Variation of the Surface redshift of the wormhole with respect to $r$.}
\end{figure*}

	\begin{eqnarray}\label{eq35}
		Z_{s}&=&-1+\frac{1}{\sqrt {g_{\it tt}}} =-1+{\frac{1}{\sqrt{e^{-\frac{2\beta}{r}}}}}.
	\end{eqnarray}

We have plotted the variation of the surface redshift along the radial expanse of the wormhole in Figure 11. The surface redshift is found to be maximum at the throat and drops down as we move towards the surface of the wormhole. Throughout the wormhole, the surface redshift is less than 2, which is necessary to indicate its stability\cite{Bohmer2006}.

\subsubsection{Acceleration and nature of the wormhole}

As already discussed for the previous wormhole model, the radial component of four acceleration of a static observer which can be computed from the redshift and the shape functions of the wormhole, determines whether the wormhole has an attractive or repulsive geometrical nature, the former implying the necessity of an outward-directed radial acceleration to prevent the observer from getting pulled into the wormhole, while the latter implying the necessity of an inward-directed radial acceleration to prevent the observer getting pushed away from it. The radial component of the four-acceleration for the present wormhole can be computed to have the form

\begin{eqnarray}\label{v5}
	a^r=-{\frac {8\beta{v}^{2}}{r\rho_{{c}}\left( 2\beta+r \right)}\left(\sqrt{\rho_{0}C_1}r^2 \left(\omega+
\frac{1}{2}\right){{\rm e}^{{\frac{(\mu+1)\beta}{2\mu r}}}}+2{{C_1}}^{2} r^2  {\rho_{0}}^{2}\pi  \left( \mu+\frac{1}{2} \right)
\left( {{\rm e}^{{\frac{(\mu+1)\beta}{\mu r}}}}\right)^{2}-\pi{{\rm e}^{{\frac{(\mu+1)\beta}{\mu r}}}}C_1\mu r^2\rho_{0}\rho_{{c}}-\frac{1}{8}\rho_{{c}}\right)}
\end{eqnarray}
	
\begin{figure*}[thbp] \label{Radacc}
	\centering
	\includegraphics[width=0.5\textwidth]{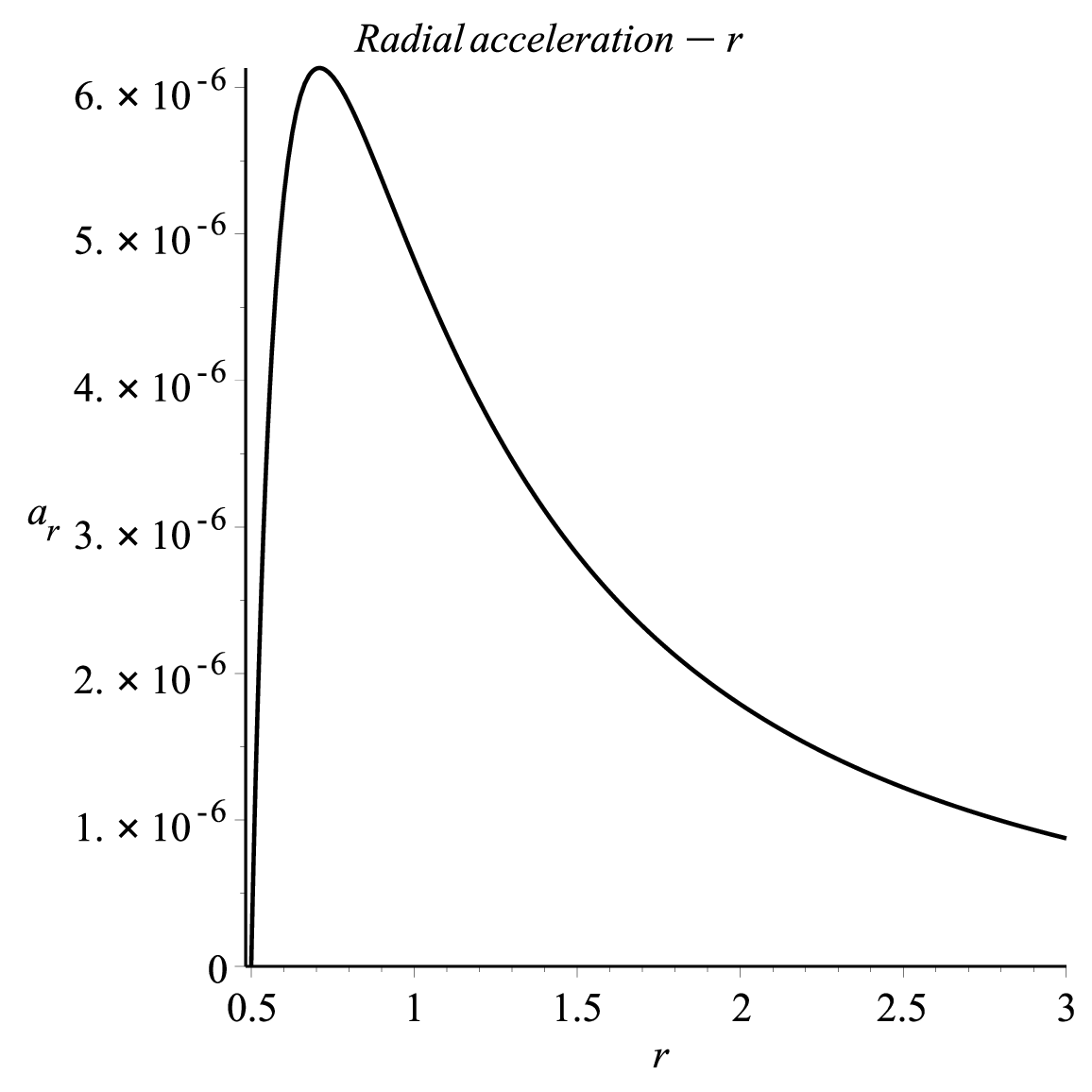}
	\caption{Plot of radial component of acceleration with respect to $r$.}
\end{figure*}
We have plotted the variation of the radial acceleration along the radial distance r in Figure 12. It turns out that the radial acceleration is positive all along as we approach the surface of the wormhole from the throat, thus establishing an attractive geometrical nature, as opposed to the Kuchowicz wormhole. 	

\textbf{{The NEC violated here is different from the violation of NEC in relativistic wormholes as in the later case it is violated by sourcing the wormhole with exotic matter or introduction of additional fields, but for both the wormhole models constructed on the brane, there is no necessity of  introduction of any exotic matter or additional fields or $\Lambda$-term (cosmological constant) to violate the NEC. The NEC is violated purely due to the high energy (UV) correction terms introduced from the brane and bulk effects in the form of the quadratic stress energy and projected bulk Weyl tensor on the brane which acts effectively as stress energy of some additional matter on the brane.}}.
	
	\section{DISCUSSIONS AND CONCLUSION}
	
	In this paper we have attempted to construct a static, spherically symmetric and  traversable wormhole on the Sahni-Shtanov braneworld with a timelike extra dimension. It is well known that in a standard relativistic context, a wormhole structure connecting two different spacetime points is unstable at the throat due to the development of a Weyl curvature singularity as a consequence of infinitely large tidal forces. So the wormhole can not act as a bridge facilitating a shortcut path reducing travel time and distance between the two spacetime points as speculated by Einstein. A solution to this problem can be approached bidirectionally, either by modifying the geometry sector via the modification of the EH action or modifying the matter source Lagrangian by introducing sources that violate one or more of the energy conditions. The first solution was provided by Morris and Thorne suggesting the introduction of some type of matter at the throat capable of violating the NEC known as exotic matter. Such matter being characterized by negative pressure is gravitationally repulsive in nature. However, a possible way to avoid the use of such exotic matter violating the NEC is to modify the gravity sector as already discussed. Braneworld gravity provides an ideal framework for doing this as the gravitational contributions coming from the higher dimensional bulk can significantly modify the gravitational dynamics on the brane at considerably high energies. It is to be mentioned here that the braneworld scenario can also significantly modify the gravitational dynamics at low energies(late times)~\cite{DDG,Sahni2} with a spacelike extra dimension considering a non-vanishing curvature term on the brane($m\neq 0$) and for a non-zero brane tension($\sigma\neq 0$)~\cite{Sahni2}, both the SEC as well as NEC is effectively violated on the brane leading the universe to accelerate at late times with an effective phantom behaviour ($\omega_{eff}<-1$). As there is no requirement for an exotic phantom fluid to exist physically, so the asscoiated problems with ghosts and instablities appearing in the relativistic context can be avoided and the acceleration is a consequence of the IR corrections to standard GR introduced with a spacelike extra dimension and the induced curvature on the brane.
	
	\textbf{{A supernegative EoS ($\omega<-1$) gives rise to some new kinds of singularities. These singularities are not of a conventional nature as the big bang or crunch singularities and have been studied in some details in the literature\cite{O1,O2,O3,SS1,SS2,PS,O4,O5}. These singularities were first classified into four types by Nojiri, Odintsov and Tsujikawa\cite{OS1} in there seminal work. The previously known big bang and crunch singularities are characterized by vanishing scale factor and diverging energy density along with Riemann curvature. Similarly, the black hole singularity is also characterized by diverging energy density and Riemann curvature. However, at the wormhole throat the energy density does not diverge and it is the tidal force and the Weyl curvature that diverges. So, it is more interesting to talk about these new future singularities occurring in a finite lifetime of the universe as some of them may be characterized by finite energy densities just like the singularity appearing in a wormhole throat. Moreover, the energy conditions are satisfied in the big bang, crunch, or black hole singularity while they are violated in the wormhole just like some of these new singularities.}}
	
	\textbf{{The first type of future singularity also known as the Big Rip is accompanied a violation of the NEC. The curvature invariants diverge along with diverging scale factor, energy density and pressure. The Type II singularity is also known as the sudden singularity and is characterized by a finite scale factor and energy density but the pressure is found to diverge leading to a diverging scalar curvature and associated violation of the dominant energy condition $\rho\geq|p|$. Thye type III singularity is similar to the type II but the energy density also diverges. Thye type IV singularity is also very interesting as the energy density, pressure, scale factor and curvature all remain finite, however the derivatives of the curvature diverge. Type III and IV singularities can also be accompanied byviolation of  certain energy conditions but it depends on the model under consideration. In the braneworld scenario with $\epsilon=1$ and $m\neq0$, where there is effective phantom behaviour $\omega_{eff}<-1$, such quiescent past and future singularities are known to occur during the contracting and expanding phases, respectively, resulting from the brane being embedded in the higher dimensional bulk spacetime\cite{SS1}. These singularities are also peculiar in the sense that they are characterized by finite density, pressure, scale factor and Hubble parameter but the higher derivatives of scale factor along with the curvature invariants are found to diverge.}}
	
	\textbf{{However, the evolution of the universe with a supernegative equation of state is itself peculiar since the energy density of matter grows increasingly with the expansion resulting in increased significance of quantum corrections before the singularity is approached. It was first shown by Nojiri and Odintsov~\cite{O2} that the involvement of such quantum effects close to the singularity in the form of back reaction of conformal quantum fields can possibly dilute the singularity resulting in an asymptotically deSitter phase. They also subsequently generalized their analysis to include the braneworld scenario. A similar approach to the IR corrected braneworld scenario with effective phantom behaviour also results in a milder singularity, where derivatives of the Hubble parameter up to the second order remain finite and third order derivatives diverge. Additional inclusion of quantum effects via particle production of non-conformal fields may result in an oscillating Hubble parameter~\cite{SS2}. Further, it has been shown by Nojiri and Odintsov~\cite{O3} that the thermodynamical consequences of the diluted singularity lead to well defined quantum corrected entropy bounds near it. An extensive analysis of past and future singularities in the framework of LQC can be found in~\cite{PS,O4}. A bouncing cosmology with type IV singularity in F(R) gravity has been studied in\cite{O5}.}}	
		
	The braneworld framework we have used in this paper with a timelike extra dimension is a dual to the Randall-Sundrum single brane model that modifies the dynamics on the brane even more drastically. In the RS model there is a correction term quadratic in stress-energy appearing with a positive sign due to the Lorentzian signature of the bulk. On the contrary, in the SS model, due to the bulk signature deviating from being Lorentzian, a negative sign appears before the quadratic correction term. This results in the avoidance of both the initial singularity and any possible future big crunch singularity provided the matter source satisfies $\rho+p\geq 0$, which is just the opposite condition required in a standard relativistic context for any likelihood of avoiding the singularity. If the matter source also obeys the condition $\rho+3p>0$, then the universe goes through an infinite number of non-singular bounces avoiding both the big bang and big crunch resulting in a cyclic universe. Additionally, the effective energy density and effective radial and tangential pressure terms on the brane appearing due to the projection of the bulk Weyl tensor on it also appear with the opposite sign due to the alteration of the bulk signature. So, it would be really interesting to check whether a static, spherically symmetric and traversable wormhole without assuming any form of charge to be possessed by the matter distribution, can be constructed on the SS brane and if so, under what conditions. Also, we study a few features of such a wormhole.
	
	We have assumed the redshift function to be described by a Kuchowicz metric potential. The physical motivation behind the use of this function comes from the fact that it has been used to model the interior of gravastars where the central curvature singularity of a black hole is removed by assuming an ad hoc EoS source by gravitational condensate matter. Since, for the wormhole throat to be stable and traversable, the Weyl curvature singularity is to be avoided at the throat, so using a regular well behaved metric function that remains finite for finite radial distances may be a good choice. Moreover, such a metric potential has also been used as a wormhole redshift function in literature~\cite{Sengupta2}. With the assumed redshift function and assuming perfect fluid matter on the brane described by a linear EoS, we use the modified EFE on the SS brane along with the stress-energy conservation equation to obtain the energy density of matter in the wormhole and the unknown metric potential in the form of the wormhole shape function. As we can see from the variation of the energy density along the radial distance in Fig. 1, the energy density remains positive throughout the wormhole and it is minimum at the throat but rises exponentially as we move towards the surface of the wormhole. It might appear apparently that a non-vanishing energy density at the surface violates the flare-out condition and the asymptotic flatness, but the effective energy density vanishes at the surface due to the higher dimensional corrections in energy and thus asymptotic flatness and flaring out is ensured. It is the effective matter distribution on the brane which includes the normal physical matter plus the effective matter arising due to local and non-local corrections that makes the construction of the traversable wormhole possible. This shows that there is no requirement for exotic matter to stabilize the throat as any such requirement would have maximized the matter distribution at the throat rather than minimizing it. The obtained shape function is also plotted along the radial distance and its shape resembles the shape of the wormhole very aptly. It is to be noted that the other desirable properties that must be possessed by a shape function to describe a physically consistent wormhole model, as prescribed by Morris and Thorne are also satisfied. Namely, (i) $b(r_0)=r_0$, (ii)  $\frac{b(r)}{r}<1$ for all $r>r_0$, where $r_0$ denotes the throat radius. The bending of the shape function for an increasing radial distance from the throat indicates the presence of a potential required to be overcome by the traveller which justifies the repulsive geometry of the wormhole.
	
	Next, the validity of the NEC has to be verified for matter constituting the wormhole on the brane. It is an essential condition that the NEC must be violated in order to ensure traversability and stability of the wormhole. As evident from Fig. 3, it turns out that the NEC is violated for an effective matter description on the brane, although we consider normal matter constituting the wormhole on the brane. This means that the sum of the effective density and effective pressure on the brane turns out to be negative for both the radial and tangential components. For drawing all the plots, we do not choose any arbitrary value for the constant model parameters, but as as shown in the section on junction conditions, we have obtained the parameters using  the boundary conditions. First, the surface density and surface pressure of the wormhole is obtained making use of the Israel-Darmois junction conditions. The boundary conditions are then obtained making use of the junction conditions. Physically justifiable values are chosen for some of the model parameters, namely the throat radius $r_0$, the wormhole surface at $r=R$, the mass of the wormhole $M$, the tidal charge $Q$, the Kuchowicz parameter $C$ and the critical density $\rho_c$. Applying these values and using the junction conditions, we evaluate the other unknown parameters including the Kuchowicz parameter $B$, the EoS parameters $\mu$ and $\omega$, the unknown constant in the bulk contribution to the effective energy density $\rho_0$ and the integration constants $C_1$ and $C_2$.
	
	It turns out that for positive values of the EoS parameter implying gravitationally attractive matter, the NEC is still violated on the brane by the effective matter due to the presence of extra dimensional effects and the obtained shape function satisfies all the necessary criteria mentioned above. This can be accounted for by the high energy or UV corrections to gravity arising from the SS braneworld, where the local corrections arise as quadratic terms in the effective stress-energy tensor components while the non-local terms arising from the projection of the bulk Weyl tensor on the brane may be interpreted as some additional matter with its own effective energy density and effective pressures varying in the radial and tangential components that have an effective radiation-like behaviour. As already discussed earlier, an additional consequence of this Weyl projection is the presence of a tidal charge of positive magnitude in the vacuum spacetime exterior to the wormhole surface on the brane. This also contributes to the fact that normal matter can constitute a wormhole on the SS brane due to the effective violation of NEC. On varying the assumed model parameters, we find that a stable traversable wormhole can be obtained on the brane for multiple positive values of the EoS parameter $\mu$ but the best obtained wormhole shape is found at a value of 0.41. The corresponding value of the other EoS parameter describing the effective matter responsible for bulk contribution to the brane turns out to be around 0.345. Importantly, the Kuchowicz parameter $B$ turns out to be negative, which for a unit value of the other Kuchowicz parameter $C$ guarantees the asymptotic flatness as both the redshift and shape functions turn out to be finite as we move to infinitely large radial distances away from the throat of the wormhole.
	
	Another important feature to ensure the traversability of a wormhole is the finiteness of the tidal acceleration at the throat. The radial and tangential tidal accelerations are expressed in terms of components of the Riemann curvature tensor. We constrain the velocity of a traveller traversing the wormhole throat by putting a realistic limit on the tangential tidal acceleration. Also, the radial acceleration is sufficiently small to ensure that there is no possibility of occurrence of any Weyl singularity at the throat that can rip the traveller apart while attempting to traverse the throat. The Weyl singularity is avoided at the throat with matter obeying the NEC due to the extra dimensional effects of gravity coming into play. As gravity is free to access the bulk spacetime, the extra dimensional effect can be realized on the brane gravitationally. The effective matter in the two different spacetimes connected by the wormhole throat on the SS brane do not gravitationally attract the throat in opposite directions making it unstable. This can be further verified using a linearized stability analysis, using which we obtain the regions of stability of the wormhole in Fig. 4 denoted by regions 1,2 and 3 in terms of the introduced parameter $\zeta$. The parameter $\zeta$ is evaluated in the effective potential formalism where the potential is constructed from the surface density and $\zeta$ is obtained in terms of the surface density and surface pressure. For the equation of motion we obtained on making the throat radius to be a function of proper time, stable regions indicate a minima in the effective potential which is essentially the first order derivative of the potential with respect to the throat radius being zero and the second order derivative being a positive quantity and this is mathematically reduced to an inequality in terms of the parameter $\zeta$. We also obtain the radial four-acceleration of the wormhole and plot its variation along the radial distance in Fig. 5 in order to assess the attractive or repulsive nature of the wormhole geometry obtained on the SS brane. If the quantity turns out to be negative then the wormhole geometry is repulsive, meaning that a static observer in the vicinity of the wormhole mouth shall require an acceleration in the radially inward direction not to be pushed away from the wormhole. Our wormhole model on the SS brane is found to be characterized by such a repulsive geometry. So, an inward directed radial acceleration is required by the observer to traverse the wormhole. The surface redshift has been computed for our wormhole model and the obtained values on variation of the radial distance also indicate towards the stability of our wormhole model.
	
	\textbf{{Since, there is ample freedom for choice of matter on the brane, the wormhole constructed in model 1 is not an unique one. There can exist other wormholes on the brane for different choice of the redshift function. In order to demonstrate this, we have chosen a second redshift function in model 2 and investigated whether a stable, traversable Lorentzian wormhole can be constructed out of it. Following our analysis in model 1, we obtain the energy density and effective energy density for matter constituting the wormhole. Unlike model 1, the density decreases away from the throat as we move radially towards the surface but this is not sufficient to ensure flare-out of the throat or asymptotic flatness of the wormhole. This is rather ensured by the reduced effective energy density near the surface. The NEC is violated by the effective matter distribution and so ordinary matter can be used to construct wormholes on the brane due to brane and bulk corrections in the modified field equations. The junction conditions have been computed at the wormhole surface by computing the surface energy density and pressure arising due to the extrinsic discontinuity at the surface which behaves as a junction between the interior and exterior spacetimes. The vanishing of the surface density and surface pressure at the wormhole surface along with the continuity of the metric potential and its derivatives along the junction are used to compute the unknown model parameters. The values of these computed model parameters have been used for plotting the graphs. The tidal acceleration in both the radial and tangential directions are of the order of acceleration due to gravity on the earth ensuring safe passage of a traveller through the wormhole. The later condition of limiting the tangential tidal acceleration leads to constraining the upper limit of the velocity of the traveller to a realistic one.This ensures the traversability of the wormhole on the brane. A two fold stability analysis is performed like in the case of model 1, where the regions of stability have been plotted by performing a linearized stability analysis using an effective potential formalism as discuused earlier. Also, the surface redshift is computed and found to be well within desirable limits. The geometric nature of the wormhole is found to be attractive implying necessity of an outward directed radial acceleration to prevent a static observer being pulled into it. Thus, stable traversable wormhole models can be constructed for different geometries on the brane owing to the independence of choice of matter on the brane.}}.
	
	It is to be noted that for our wormhole model both the SEC and NEC are respected by the matter constituting the wormhole on the brane. In braneworld model with a timelike extra dimension, a singularity free universe can be realized due to the avoidance of the initial big bang singularity and possible future big crunch singularity such that the singularity is replaced by a bounce, provided the constituent matter obeys the NEC. For our wormhole model also if the matter constituting the wormhole obeys the NEC a stable wormhole can be obtained. Also, in such a braneworld model there can be an infinite number of bounces leading to a cyclic universe with each successive cycle characterized by an increase in amplitude leading to a natural resolution of the flatness problem in standard cosmology provided the constiuent matter obeys SEC. Our wormhole model is in agreement with this condition too, as again for the wormhole to be stable and divergence free at the throat by avoiding infintely large tidal forces, the constituent matter of the wormhole on the brane must obey the SEC. So, for obtaining a traversable wormhole on the SS brane, both the conditions necessary to realize a non-singular bounce and a cyclic cosmology must be obeyed in the absence of an induced curvature term on the brane. This leads to the realization that wormholes can be expected to exist naturally in such a higher dimensional cyclic universe on the brane. The tubular structure of the wormhole spreading out at infinitely large distances to be asymptotically flat is realized much better than in braneworld with a spacelike extra dimension and also wormhole can be constructed with considerably lower mass for a timelike extra dimension, thus minimizing the amount of matter required. One finds that SS braneworld model is capable of realizing singularity free solutions in both cosmological context and in the context of a wormhole with matter source obeying the energy conditions. Both the Riemann curvature singularity due to diverging energy density and the Weyl curvature singularity due to diverging tidal forces at the wormhole throat can be resolved in the SS braneworld model with normal matter. Moreover, this extra-dimensional model is capable of resolving some of the shortcomings of standard GR at the UV limit. A possible reason for this may be that due to the presence of the timelike extra dimension, some gravitationally repulsive effect takes over (\textit{which is not due to the presence of any effective $\Lambda$ or electrodynamic charge and depends solely on the extra dimension being timelike}) when the energy density or tidal force grows extremely large and this effect is responsible for turning the singular collapse, inevitable in the context of standard GR, into a smooth non-singular transition where the usual notion of spacetime is preserved.

	\section*{Acknowledgments}
	
	The authors are extremely thankful to Prof. Varun Sahni (IUCAA, Pune) and Prof. Yuri Shtanov (BITP, Kiev) for their kind and extremely helpful discussions and comments without which the paper would not have been possible in the present form. The authors are also grateful to them as the writing of the paper is largely inspired from the discussions with them.
	
	MK is thankful to the Inter-University Centre for Astronomy and Astrophysics (IUCAA), Pune, India for providing the Visiting Associateship under which a part of this work was carried out. RS is thankful to the Govt. of West Bengal for financial support through SVMCM scheme. SG is thankful to the Directorate of Legal Metrology under the Department of Consumer Affairs, West Bengal for their support.

	~~~~~~~~~~~~~~~~~~~~~~~~~~~~~~~~~~~~~~~~~~~~~~~~~~~~~~~~~~~~~~~~~~~~~~~~~~~~~~~~~~~~~~~~~~~~~~~

\end{document}